\documentclass[
    amssymb, amsmath,
    floatfix,
    reprint,%
    twocolumn,%
  ]{revtex4-1}

\usepackage[colorlinks=true,linkcolor=blue]{hyperref}
\usepackage{graphicx}% Include figure files

\usepackage[T1]{fontenc} % Use modern font encodings
\usepackage[utf8]{inputenc}
\usepackage{lmodern}

\usepackage{booktabs}
\usepackage{subcaption}
\usepackage{etoolbox}
\usepackage[separate-uncertainty=true]{siunitx}
\DeclareSIUnit{\calorie}{cal}
\DeclareSIUnit{\kcal}{\kilo\calorie\per\mol}
\DeclareSIUnit\Molar{\textsc{m}}
\usepackage{upgreek}
\usepackage{chemmacros}

\chemsetup{
 formula = mhchem,
 greek = chemgreek,
 modules = thermodynamics,
 modules = redox,
 modules = reactions
}

\renewrobustcmd{\bfseries}{\fontseries{b}\selectfont}
\renewrobustcmd{\boldmath}{}

\newcommand{\etal}{et al.}

\begin{document}
\title{Properties of the Tetravalent Actinide Series in Aqueous Phase from Microscopic Simulation: a Self-Consistent Engine}
% Use letters for affiliations, numbers to show equal authorship (if applicable) and to indicate the corresponding author
\author{Eléonor Acher}
\affiliation{CEA, Nuclear Energy Division, Research Department on Mining and fuel Recycling Processes (LILA), BP17171 F-30207 Bagnols-sur-Cèze, France}
\altaffiliation{Univ. Lille, CNRS, UMR 8523 – PhLAM – Physique des Lasers Atomes et Molécules, F-59000 Lille, France}
\author{Michel Masella}
\affiliation{Laboratoire de Biologie Structurale et Radiobiologie, Service de Bioénergétique, Biologie Structurale et Mécanismes, Institut Joliot, CEA Saclay, F-91191 Gif sur Yvette Cedex, France}
\author{Valérie Vallet}
\affiliation{Univ. Lille, CNRS, UMR 8523 - PhLAM - Physique des Lasers Atomes et Molécules, F-59000 Lille, France}
\author{Florent Réal}
\affiliation{Univ. Lille, CNRS, UMR 8523 - PhLAM - Physique des Lasers Atomes et Molécules, F-59000 Lille, France}
\email{florent.real@univ-lille.fr}

\begin{abstract}
In the context of nuclear fuel recycling and environmental issues, the understanding of the properties of radio-elements with various approaches remains a challenge regarding their dangerousness. Moreover, experimentally some issues are also of importance; first, it is imperative to work at sufficiently high concentrations to reach the sensitivities of the analytical tools, however this condition often leads to precipitation for some of them, second stabilizing specific oxidation states of some actinides remains a challenge, thus making it difficult to extract general trends across the actinide series. Complementary to experiments, modeling can be used to unbiasedly probe the actinide's properties in aquatic environment and offers a predictive tool. We report the first molecular dynamics simulations based on homogeneously built force fields for the whole series of the tetravalent actinides in aqueous phase from \ce{Th^{IV}} to \ce{Bk^{IV}} and including \ce{Pu^{IV}}. The force fields used to model the interactions among the constituents include polarization and charge donation microscopic effects. They are built from a self-consistent iterative \textit{ab initio} based engine that can be included in future developments as an element of a potential machine learning procedure devoted to generate accurate force fields.
The comparison of our simulated hydrated actinide properties to available experimental data show the model robustness and the relevance of our parameter assignment  engine. Moreover our simulated structural, dynamical and evolution of the hydration free energy data show that, apart from \ce{Am^{IV}} and \ce{Cm^{IV}}, the actinide properties change progressively along the series.
\end{abstract}

\maketitle

%Please use \dag to cite the ESI in the main text of the article.
%If you article does not have ESI please remove the the \dag symbol from the title and the footnotetext below.
%additional addresses can be cited as above using the lower-case letters, c, d, e... If all authors are from the same address, no letter is required
%%%%%%%%%%%%%%%%%%%%%%%%%%%%%%%%%%%%%%%%%%%%%%%%%%%%%%%%%%%%%%%%%%%%%
%% The "tocentry" environment can be used to create an entry for the
%% graphical table of contents. It is given here as some journals
%% require that it is printed as part of the abstract page. It will
%% be automatically moved as appropriate.
%%%%%%%%%%%%%%%%%%%%%%%%%%%%%%%%%%%%%%%%%%%%%%%%%%%%%%%%%%%%%%%%%%%%%
%\begin{tocentry}
%\includegraphics[width=\linewidth]{Figure-TOC.pdf}
%\end{tocentry}

Because of its importance to separation science and to modeling actinide migration from nuclear waste repositories, there is a need of a profound knowledge of actinide environmental behavior. This involves a predictive understanding of actinide solution chemistry as the first critical step~\cite{actinide-Knope-CR2013-113-944}. Namely, actinide environmental behavior implies speciation in the bulk phase but also at mineral-aqueous solution interfaces. For the later, sorption reactions can either involve pure electrostatic sticking on a mineral surface with little change occurring in the first metal ion hydration shell, or inner-sphere surface complexation with chemical binding between the actinide and the mineral surface functional groups. Beside concentration, pH, and redox conditions, the competition between these two mechanisms is controlled by the stability of the cation first hydration shell, namely the hydration energy, as it governs its propensity to form strong inner sphere binding~\cite{actinide-Geckeis-CR2013-113-1016}.

In this context, the behavior of tetravalent actinides, \ce{An^{IV}}, in solution is highly relevant for two application fields connected to the nuclear industry : their migration as mentioned previously but also their separation from the spent nuclear fuels~\cite{actinide-Nash-An-Tn-Book2008-2622}. However, the radiotoxicity and the complex chemistry of these elements explain the rather scarce availability of experimental data for them, in particular in aqueous phase. Whilst extended X-ray absorption fine structure (EXAFS) technique yields highly accurate \ce{An^{IV}}/water oxygen  bond lengths in the series \ce{Th^{4+}}-\ce{Bk^{4+}}),  there are still large uncertainties in the experimental  \ce{An^{IV}} hydration numbers, $N_c$, ranging from eight to thirteen for \ce{Th^{4+}}-\ce{Bk^{4+}}~\cite{actinide-Moll-IC1999-38-1795,actinide-Torapava-IC2009-48-11712,actinide-Hennig-IC2007-46-5882,actinide-Neck-RA2002-90-485, actinide-Wilson-ACIE2007-46-8043,actinide-Banik-DT2016-45-453,actinide-Moll-IC1999-38-1795,actinide-Ikeda-Ohno-IC2009-48-7201,actinide-Hennig-IC2005-44-6655,actinide-Antonio-RA2001-89-17,actinide-Ikeda-Ohno-IC2008-47-8294,Denecke-Tal2005-65-1008,actinide-Rothe-IC2004-43-4708,actinide-Dardenne-RA2009-97-91,actinide-Antonio-RA2002-90-851,lanthanide-Sham-PRB1989-40-6045}.

To complement experimental data, research groups have either developed empirical theoretical models or performed quantum chemical computations on small \ce{An^{IV}}/ligand clusters. We may refer here to the Born-like model (adjusted from geometrical experimental data) of David~\cite{actinide-David-JNM1985-130-273,actinide-David-JLCM1986-121-27} that yields \ce{An^{IV}}  hydration free energies, $\Delta G_\mathrm{hyd}$ values. These values decrease regularly along the \ce{An^{IV}} series. In 2003, David and Vokhmin~\cite{actinide-David-NJC2003-27-1627} revised the  approach. They got then a very close tendency, at the remarkable exception of \ce{Np^{IV}} whose $\Delta G_\mathrm{hyd}$ value is down-shifted by more than \SI{25}{\kcal} with respect to the 1986 trend. A similar smoothly decreasing trend, although less steep, was also obtained by Bratsch and Lagowski~\cite{lanthanide-Bratsch-JPC1985-89-3310,lanthanide-Bratsch-JPC1985-89-3317,actinide-Bratsch-JPC1986-90-307} in 1986 with their simple model based on ionic radii. We may also cite the study of Banik~{\etal}~\cite{actinide-Banik-DT2016-45-453} who performed quantum calculations on a set of small \ce{An^{IV}} hydrated clusters (with bulk effects accounted for using a polarizable continuum solvent approach). These authors got \ce{An-O_{water}} bond lengths in line with EXAFS data, and they concluded to a decrease of the \ce{An^{4+}} hydration number from nine to eight, with a turning point at \ce{Cm^{4+}}.  

To go beyond static models, molecular dynamic, MD, simulations have also been considered. However their predictive capability relies on the accuracy of the potential energy used to model microscopic interactions. Potential energies computed from quantum chemistry methods are the most relevant to investigate the properties of any kind of microscopic systems. Regarding \ce{An^{IV}}, we may cite MD studies focusing on the hydration process of  \ce{Th^{4+}}, \ce{U^{4+}}, and \ce{Pu^{4+}} using density functional theory (DFT) methods~\cite{actinide-Spezia-JPC2012-116-6465,actinide-Atta-Fynn-IC2012-51-3016,actinide-Odoh-JPC2013-117-12256,actinide-Spezia-PCCP2014-16-5824,actinide-Atta-Fynn-JPC2016-120-10216}. However none of the available DFT methods is able to describe accurately both water/water and cation/water interactions, at least when comparing DFT results to those of higher level quantum \textit{ab initio} methods like the M{\o}ller-Plesset second-order perturbation theory, MP2, and above~\cite{actinide-Real-JPC2010-114-15913} (see discussions of Section~1 of the Supplementary Material). As high end quantum methods cannot be still considered to efficiently simulate bulk systems, even when used in hybrid quantum mechanical/molecular modeling schemes~\cite{water-Zen-JCP2015-142-144111,water-Liu-CS2018-9-2065}, molecular modeling approaches based on empirical interaction potentials, force fields, are still the most suited to simulate such systems on significantly long simulation times, and complex solution mixtures of different ions and different solvents.

As discussed earlier by other authors~\cite{actinide-Hemmingsen-JPC2000-104-4095,actinide-Clavaguera-Sarrio-JPC2003-107-3051,actinide-Real-JPC2010-114-15913,actinide-Real-JCP2013-139-114502},  \ce{An^{IV}}/water microscopic interactions result from a complex interplay between mainly large electrostatic, non-additive polarization and charge-donation effects whose relative magnitude is not obvious to quantify. Moreover and as far as we know, most of the water force fields proposed to date are not able to capture the magnitude of the water/water repulsive interactions in cation first hydration shells, at the exception of the TCPE/2013 one~\cite{actinide-Real-JCP2013-139-114502}. This explains the difficulty of building well balanced force fields to study the hydration of \ce{An^{IV}} and thus the very few \ce{An^{IV}} MD bulk simulations based on force fields reported to date~\cite{actinide-Real-JPC2010-114-15913,actinide-Marjolin-TCA2012-131-1,actinide-Spezia-JPC2012-116-6465,actinide-Real-JCP2013-139-114502, actinide-Real-JCC2013-34-707}. 

Besides the choice of a physically meaningful functional form for the force field, another critical issue is the strategy to assign force-field parameters. The most promising and robust strategy consists of assigning the parameters to reproduce only QM data regarding a limited set of training molecular clusters. These training data sets usually comprise small molecular clusters optimized in gas phase. We already built accordingly sophisticated force fields, hereafter name Gas-Phase Parameter (GP-P), to study the hydration of \ce{Th^{IV}} and \ce{Cm^{III}}~\cite{actinide-Real-JPC2010-114-15913,actinide-Real-JCC2013-34-707}. However such an approach implicitly assumes that all the features of bulk phase interactions can be captured by investigating gas phase systems. In the particular case of \ce{U^{IV}}, Atta-Fynn~{\etal}~\cite{actinide-Atta-Fynn-JPC2016-120-10216} showed using quantum computations cation/water oxygen distances to be shifted by about \SI{0.05}{\AA} from gas phase to aqueous solution. Because of the +4 charge of \ce{An^{IV}} elements, such a distance shift yields a change in the Coulombic interaction energy between \ce{An^{IV}} and each water oxygen of its first hydration shell amounting to about 20 k$_\mathrm{B}$T at ambient conditions. Hence a natural improvement, but more costly in term of computational ressources, would be to consider bulk phase cluster structures for building accurate force field. We will denote this kind of force-field Bulk-Phase Parameter (BP-P). The generation of an increasingly larger set of reference data ensure the robustness of the force field, and opens up the possibility to handle complex mixtures of ions and solvents.

The most common parameter fitting process is a single step procedure consisting in best reproducing the properties of a usually restraint training molecular cluster set. The ongoing increase of the available computational resources yields research groups to propose more and more sophisticated fitting procedures  where new QM target data are generated along the procedure cycles (see Figure~\ref{fig:scheme-MD-fit} as an illustration), the latter can be used in machine-learning FF fitting process. We may refer here to Monte-Carlo sampling approach of Galbis~{\etal}~\cite{actinide-Galbis-JCP2014-140-214104}, or the Force Balance method proposed by Wang~{\etal}~\cite{FF-Wang-JTCT2013-9-452}. Inspired by the available literature, we propose a coherent \textit{ab initio} model associated to a self-consistent iterative procedure to homogeneously and coherently built a class of BP-P force fields for the whole \ce{An^{IV}} series, in order to obtain metrical information on the coordination environment of the tetravalent actinide series from Th up to Bk, as well as to explore the evolution of the hydration free energies along the series, that will be confronted to the available empirical data. For comparison purpose, we also investigated the behavior of the lanthanide element \ce{Ce^{IV}} that is often considered as a non radioactive surrogate for \ce{Pu^{IV}}~\cite{actinide-Kolman-TR1999,lanthanide-Chandrasekar-SPT2019-217-62}.

\section{Methods and computational details}
\subsection{The force field}
We use the rigid water model TCPE/2013~\cite{actinide-Real-JCP2013-139-114502}. For the ion/water interactions, we consider the ab initio based force field that describes the total interaction as the sum of four energy components,
\begin{equation} 
\label{eqn:Utotal}
\Delta U=U^{\mathrm{rep}} + U^\mathrm{qq'} +U^{\mathrm{ct}} +  U^{\mathrm{pol}}.
\end{equation}
The first three correspond to repulsive $U^{rep}$, Coulombic $U^\mathrm{qq'}$ and charge-transfer $U^\mathrm{ct}$ contributions, the latter accounting for the partial "covalent" character of the actinide/water interactions. For a system of $N$ atoms, the additive terms $U^{rep}$, $U^\mathrm{qq'}$, and $U^{ct}$ are defined as:
\begin{eqnarray}
U^{\mathrm{rep}} + U^\mathrm{qq'} + U^{ct}  = && \sum_{i=1}^{N}\sum_{j, j>i}^{N}\left[A_{ij}\exp{(B_{ij}r_{ij})}+ \right.\nonumber\\
&& \left.{}\frac{q_iq_j} {4\pi\epsilon_0r_{ij}} - D_{ij}\exp{(-\frac{r_{ij}}{\beta_{ij}})}\right],
\end{eqnarray}
where $r_{ij}$ is the distance between atoms $i$ and $j$, \{$q_i$\} are the static charges located on the atomic centers, and ($A_{ij}$, $B_{ij}$, $D_{ij}$, $\beta_{ij}$) are adjustable parameters. 

Polarization effects are modeled using induced dipole moments $\mathbf{p}_i$ expressed as

\begin{equation}
\label{eqn:induceddipole}
\mathbf{p}_i = \alpha_i\cdot\left(\mathbf{E}_i^q + \sum_{j=1}^{N^*_\mu}{\mathbf{T}_{ij}\cdot\mathbf{p}_j}\right).
\end{equation}
All the $N_\mu$ nonhydrogen atoms are polarizable centers, i.e., a single point polarizability is located on each nonhydrogen atomic center. Their isotropic polarizability is $\alpha_i$. $\mathbf{E}_i^q$ is the electric field generated on the polarizable center $i$ by the surrounding static charges $q_j$, and $\mathbf{T}_{ij}$ is the dipolar interaction tensor. They both include short-range Thole's-like damping functions~\cite{md-Thole-CP1981-59-341,actinide-Real-JPC2010-114-15913,actinide-Real-JCP2013-139-114502}, with an adjustable damping parameter $\kappa$.
All the ion-water force-field parameters optimized with the GP-P and BP-P procedures described in the coming section are listed in Table S1 in the ESI.

\subsection{QM reference data for GP-P and BP-P force fields}

The QM reference data sets will be composed of total binding energies and fragment interaction energies. The first ones correspond to the chemical reaction  \ce{An^{IV} + n H2O -> An^{IV}(H2O)_n} and are thus calculated according to the equation $BE=E_{\ce{An^{IV}(H2O)_n}}-E_{\ce{An^{IV}}}-n E_{\ce{H2O}}$. This energy is then corrected from the basis set superposition error between the actinide and the water molecules following the well established counterpoise correction method with the equation:
\begin{eqnarray}
BSSE  = && E_{\ce{An^{IV}}}^{\ce{An^{IV}(H2O)_n}} - E_{\ce{An^{IV}}}^{\ce{An^{IV}}} \nonumber\\
&&+ E_{\ce{(H2O)_n}}^{\ce{An^{IV}(H2O)_n}} - E_{\ce{(H2O)_n}}^{\ce{(H2O)_n}},
\end{eqnarray}
where the subscript refers to the fragment considered and the superscript to the basis used. The corrected total binding energy is then:
\begin{eqnarray}
BE^{cpc} =&& E_{\ce{An^{IV}(H2O)_n}} - E_{\ce{An^{IV}}}^{\ce{An^{IV}(H2O)_n}} \nonumber\\
&&- E_{\ce{(H2O)_n}}^{\ce{An^{IV}(H2O)_n}} + E_{\ce{(H2O)_n}} - n E_{\ce{H2O}},
\end{eqnarray}
where the superscript has been omitted when the basis set used is the fragment's own basis set.
The fragment interaction energies are the counterpoise corrected interaction energies computed as:
\begin{equation}
E_{int}^{cpc} = E_{AB}^{AB} - E_A^{AB} - E_B^{AB},
\end{equation}
with the fragments $A$ and $B$ corresponding, on the one hand to a single water molecule of the \ce{An^{IV}(H2O)_n} cluster and on the other hand to the remaining atoms.

For all the terms defined above, using the Turbomole package (7.1 release)~\cite{prog-turbomole712} we have performed RI-MP2 method~\cite{mrpt2-Hattig-PCCP2006-8-1159,mrpt2-Hattig-JCP2000-113-5154} for the closed-shell \ce{Th^{IV}} or UMP2 for the open-shell systems (\ce{An^{IV}}, \ce{An} = \ce{Pa} to \ce{Bk}), fixing the multiplicity of the computed state at the highest value. For each gas-phase aggregates, the energy calculations were preceded by an initial geometry optimization. For the actinides, the Stuttgart-Cologne "small core" relativistic effective core potentials (60 \ce{e^-}) were used in conjunction with the associated segmented basis sets~\cite{basis-Cao-JMST2004-673-203,basis-Cao-JCP2003-118-487,ecp-Kuchle-JCP1994-100-7535}, while the augmented correlation consistent triple $zeta$ basis sets of Dunning~\cite{basis-Dunning-JCP1989-90-1007,basis-Hattig-PCCP2005-7-59}, namely the aug-cc-pVTZ, were used for the water molecules. The 1s core electrons of the oxygen as well as the 5s, 5p, 5d electrons of the actinides were not correlated in the MP2 step.
%%%%%%%%%%%%%%%%%%%%%%%%%%%%%%%%%%%%%%%%%%%%%
%%%%%%%%%%%%%%%%%%%%%%%%%%%%%%%%%%%%%%%%%%%%%

\subsection{Self-consistent BP-P force-field procedure for \ce{An^{IV}}}
%%%
%FIGURE Method 2 Scheme
\begin{figure}[ht]
%\begin{figure}
\begin{center}
\includegraphics[width=0.8\linewidth]{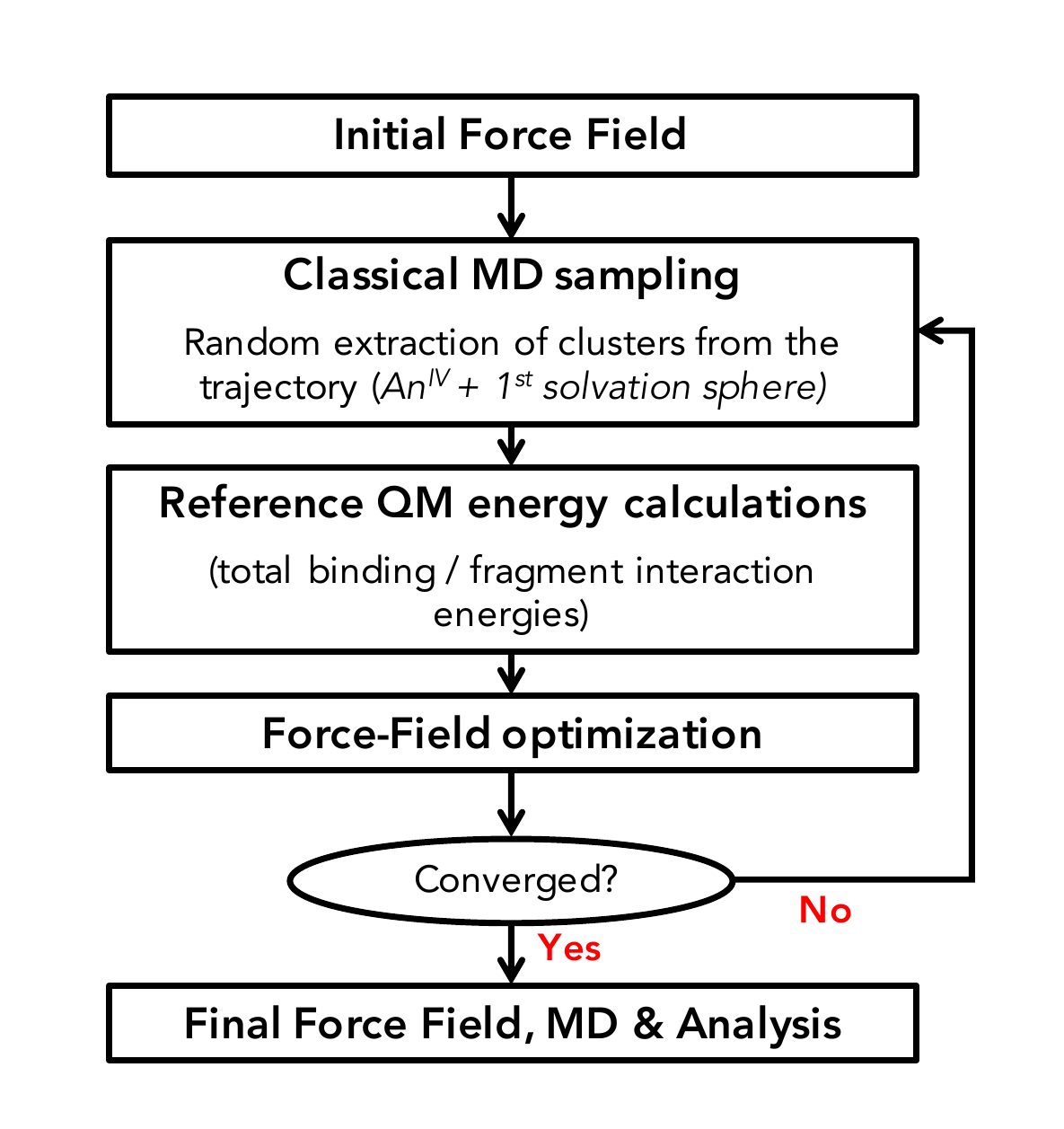}
\caption{Chart for the self-consistent iterative parameter optimization procedure. Each MD simulation is at the \SI{5}{ns} scale. The sampling consists in extracting each \SI{1}{ps} the \ce{An^{IV}} plus its first hydration shell clusters. We select then 20 of these clusters whose MM binding energies, BE, corresponding to the reaction \ce{An^{IV} + nH_2O \rightarrow An^{IV}/(H_2O)_n}, cover the range of the MM BEs corresponding to the full extracted cluster set. The reference QM computations are performed at the MP2 level.} 
\label{fig:scheme-MD-fit}
\end{center}
\end{figure}
%FIGURE Method 2 Scheme

The force-field parameters are assigned from the self-consistent iterative procedure, summarized in Figure~\ref{fig:scheme-MD-fit}. All iterative procedure iterations consist in the sequence of: MD sampling, QM reference energy calculation 
%(Binding Energies (BEs) corresponding to the reaction \ce{An^{IV} + nH_2O \rightarrow An^{IV}/(H_2O)_n} and different fragment interactions; see SI for definitions)
 and fitting process. The parameter initial guesses correspond to force-field parameters assigned from the GP-P procedure, \emph{i.e.} the training data set corresponds to small hydrated \ce{An^{IV}} clusters comprising at most 10 water molecules optimized in gas phase using  QM methods. The training data set at each iterative procedure's iteration comprises all the clusters from the previous two iterations, ensuring the BP-P convergence stability to the final force field. As high end QM computations regarding \ce{An^{IV}} are still particularly computationally expensive, the MD sampling consists in extracting 20 hydrated \ce{An^{IV}} clusters (the cation plus its first hydration sphere) from a \SI{5}{\ns} MD trajectory; for each cluster, about 10 interaction energies are computed to obtain the BEs and different fragment interaction energies. i.e 200 QM energies per cycle. To ensure that the 20~selected clusters at each iteration are representative, we checked that the MM BEs of these small clusters set span over the range of MM cluster BEs extracted from the full trajectory. 
At each iterative procedure's iteration, the evolution of the force-field parameters as well as structural properties (radial distribution function, and coordination number) is monitored, until the self-consistent solution is reached. Convergence is usually reached in less than 6 iterations, as illustrated in Figure~S2 in the ESI. Today all the computations of a single iteration of the procedure for the whole \ce{An^{IV}} series can be performed at the week-scale using a few percents of a modern supercomputing system (made of about \num{100000} computing cores).

\begin{figure}[htp]
\includegraphics[width=\linewidth]{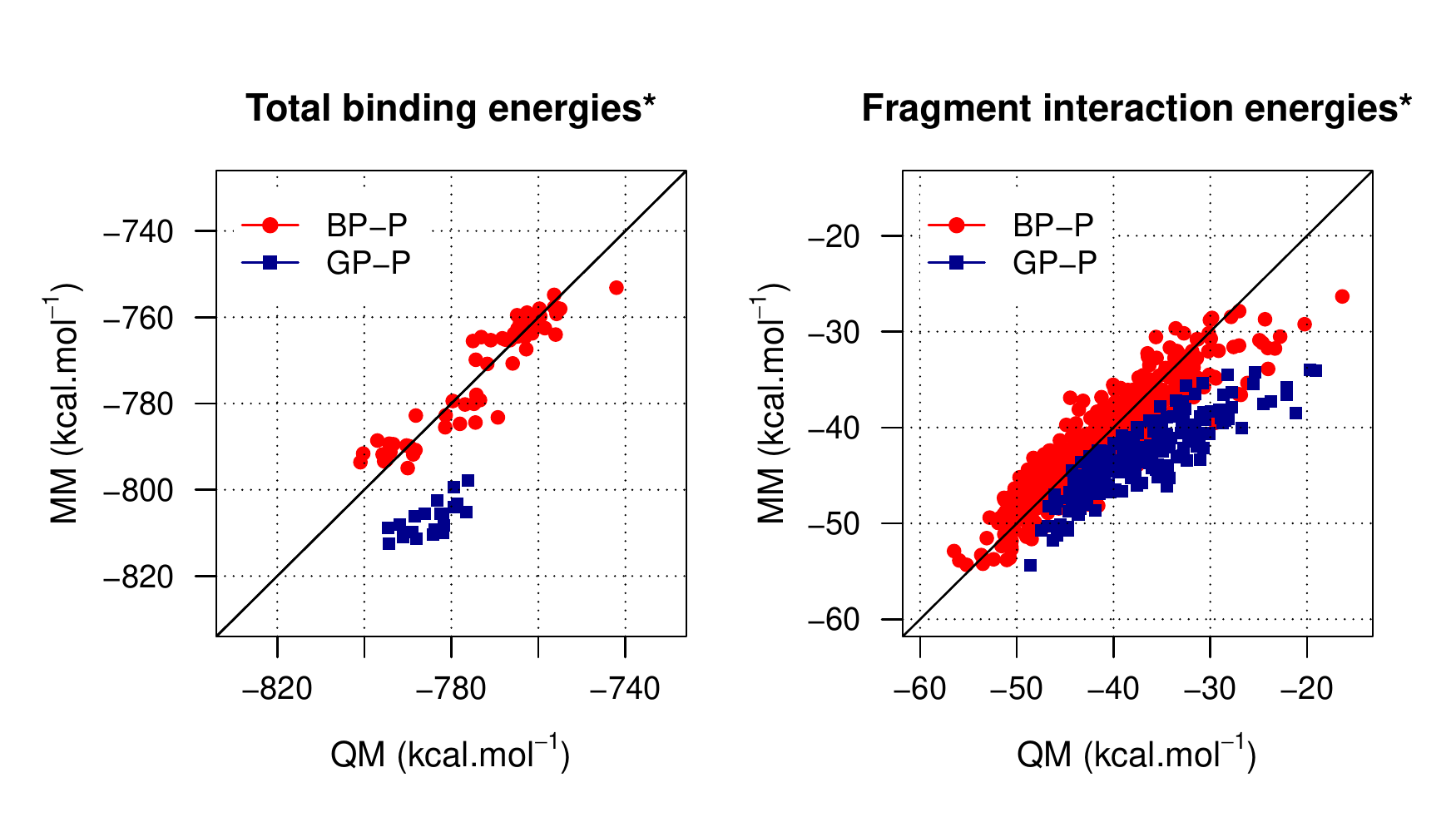}
\caption{Comparison of QM and MM binding energies BEs and of the fragment interaction energies, \emph{i.e.} the interaction energy of a single water molecule of the \ce{An^{IV}(H2O)$_n$} cluster with the remaining \ce{An^{IV}(H2O)$_{n-1}$} fragment, for \ce{Th^{IV}} when using the force-field parameter initial guess built from QM data corresponding to clusters optimized in gas phase (GP-P) and from the self-consistent iterative procedure (BP-P). The mean error is reduced to less than \SI{0.1}{\kcal} with a standard deviation of respectively 4.8 and \SI{2.6}{\kcal} for the total BEs and the fragment interaction energies with the latter approach.} 
\label{fig:QM-MM_comp}
\end{figure}

The improvement in the force-field quality arising from the iterative procedure may be assessed from the data plotted in Figure~\ref{fig:QM-MM_comp}. %In that figure, we plot the classical BEs of the training cluster set computed from the parameter initial guesses (GP-P) and from the BP-P converged force field as a function of the QM BEs. 
In this figure, the classical BEs computed from the parameter initial guesses (GP-P) and from the BP-P converged force field are plotted with respect to  their QM reference counterparts for a training cluster set. If MM BEs from the initial parameter guesses are linearly correlated with QM ones, a large systematic deviation exists between the two BE sets, of about \SI{-22}{\kcal}. If that represents only 2.5 \% of the cluster BEs, that corresponds to about 36 k$_B$T at ambient conditions. With the BP-P force field, the BE mean error is centered on 0 and its standard deviation is reduced to less than \SI{5}{\kcal}. These results support the earlier conclusion made by Tazi~{\etal}~\cite{md-Tazi-JCP2012-136-114507}, \emph{i.e.} force-field parameter adjustments relying on realistic bulk-phase structures ensures the accuracy of the resulting force field.

%
%FIGURE comparaison points QM/MM
%FIGURE comparaison points QM/MM
%

\subsection{Molecular dynamics simulations}
The MD simulations of the \ce{An^{IV}} in bulk water were conducted with a cubic box containing only one actinide cation for \num{1000} water molecules with periodic boundary conditions in PolarisMD package developed by one of us. The system is first equilibrated in volume and temperature in the \{N,P,T\} ensemble with a Noose-Hoover thermostat and barostat~\cite{md-Martyna-MP1996-87-1117}.
The different runs of production used for the parametrization are then run in the canonical ensemble~\cite{md-Liu-JCP2000-112-1685} during \SI{5}{\ns} always preceded a \SI{1}{\ns} re-equilibration. The final step of the production from which thermodynamics data and structural informations are extracted, ran for  \SI{10}{\ns}.
It should be recalled here that the water model used for the simulations is the latest TCPE/2013 version~\cite{actinide-Real-JCP2013-139-114502}. The water molecule's structural parameters constrained to their bulk equilibrium values thanks to the RATTLE algorithm (the convergence criterium is set to \SI{1e-6}{\angstrom}). Since the constituents of the system have no free internal motion, the simulation time-step is fixed to \SI{1}{\fs} and, in order to minimize the extra computational expense inherent to the polarization model, the induced dipoles are accounted for within the multiple time step r-RESPAp framework~\cite{md-Masella-MP2006-104-415}. The hydration free energies, $\Delta G_{hyd}$ are computed relative to \ce{Th^{IV}} using the standard Thermodynamic Integration (TI) scheme of twenty steps based on a linear interpolation of Hamiltonians. For each step, we performed a \SI{100}{\ps} run of equilibration followed by a \SI{500}{\ps} one during which the TI statistical average is computed. 

%\subsection*{Coordination mode analysis}
%We analyzed the coordination polyhedra using \textit{ChemNetworks}~\cite{ChemNetwoks-Ozkanlar-JCC2014-35-495}. The details of the analysis are reported Figure~S3.  

%%%%%%%%%%%%%%%%%%%%%%%
% SECTION RESULTATS & DISCUSSION
%%%%%%%%%%%%%%%%%%%%%%%

\section{Results and Discussion}

The behavior of the whole \ce{An^{IV}} series in \SI{0.03}{\Molar} cation aqueous solutions is studied by means of MD simulations at \SI{10}{ns} scale using the BP-P force fields. We do not account explicitly for counter ions in our bulk simulations. However as the simulations are performed using an Ewald summation scheme with tinfoil boundary conditions, the cations are accompanied by uniform canceling background charge, acting as an implicit dilute counter ion cloud. We also simulated accordingly the lanthanide \ce{Ce^{IV}} that is inferred to present very close hydration properties to \ce{Pu^{IV}}. From MD trajectories, relevant structural properties like the pair \ce{An^{IV}}/water oxygen radial distribution functions, RDFs, the position $d_{AnO}$ of the RDF first peak, and the cation hydration number $N_c$ are estimated. 

%FIGURE RDF_An
\begin{figure}[ht]
 \includegraphics[width=\linewidth]{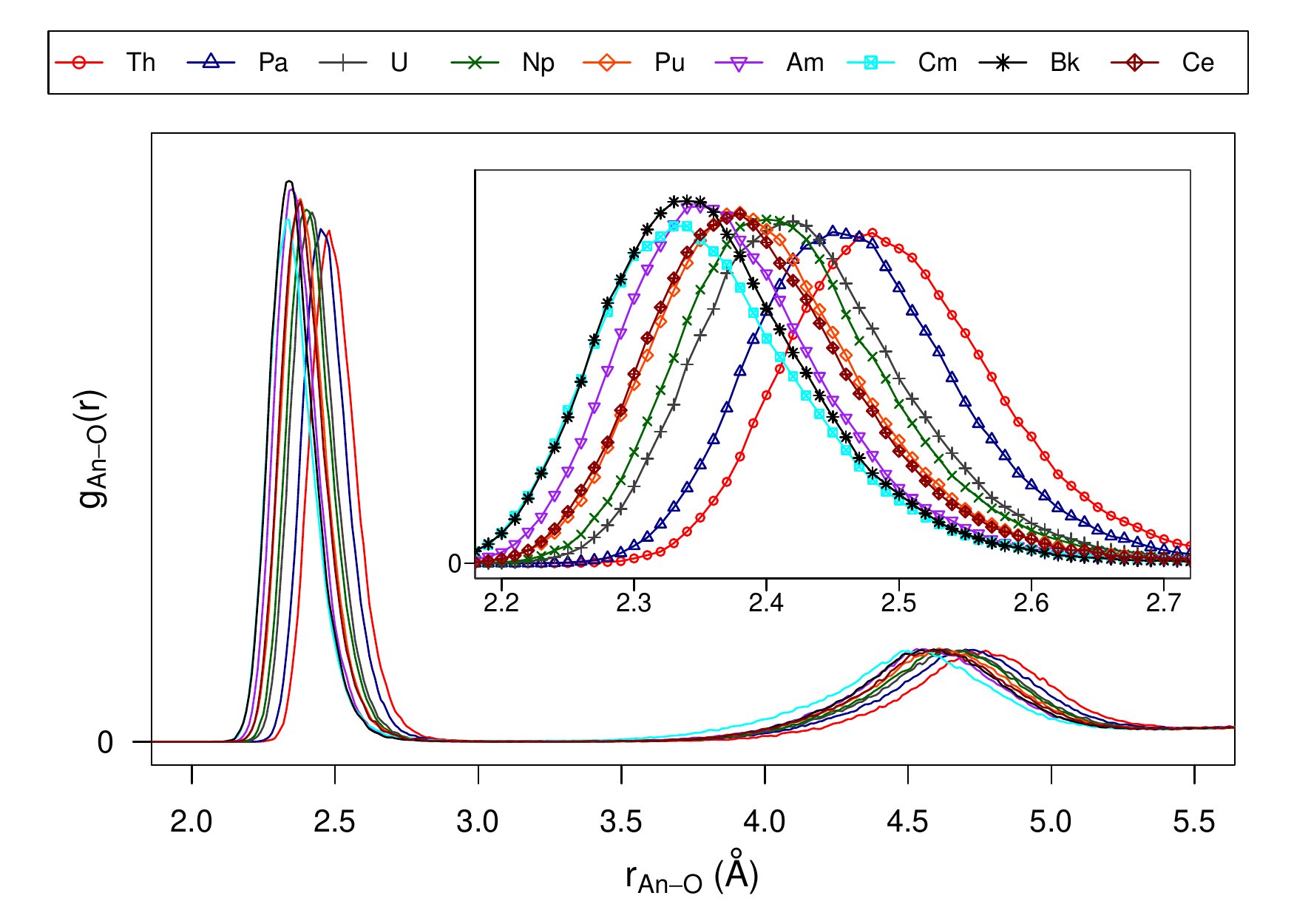}
\caption{\ce{An^{IV} - O} pair radial distribution functions. } 
\label{fig:RDF_An-O_MDfit}
\end{figure}
%FIGURE RDF_An

% RDF and d(AnO)

\begin{figure}[htp]
\includegraphics[width=\linewidth]{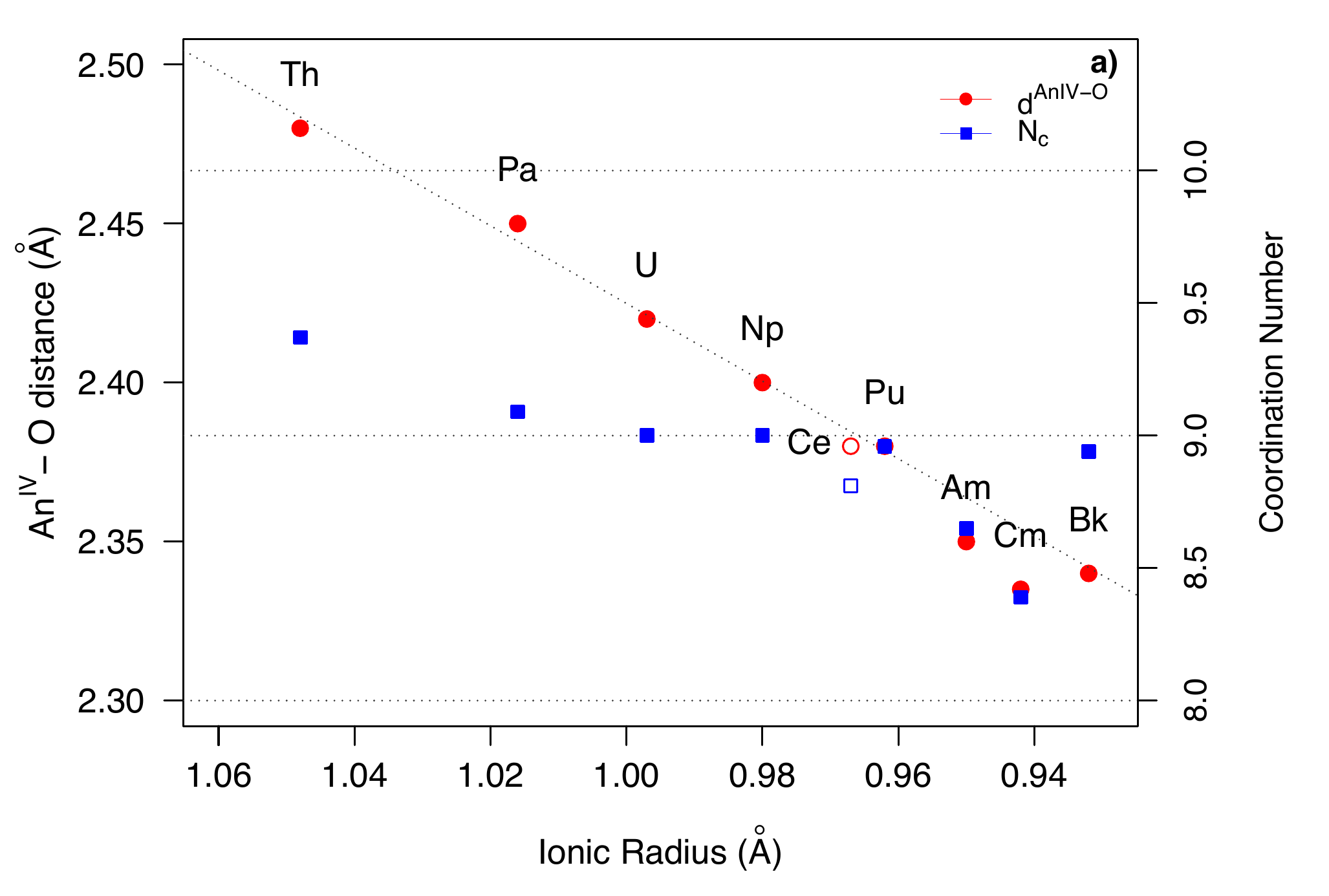}\\
\includegraphics[width=\linewidth]{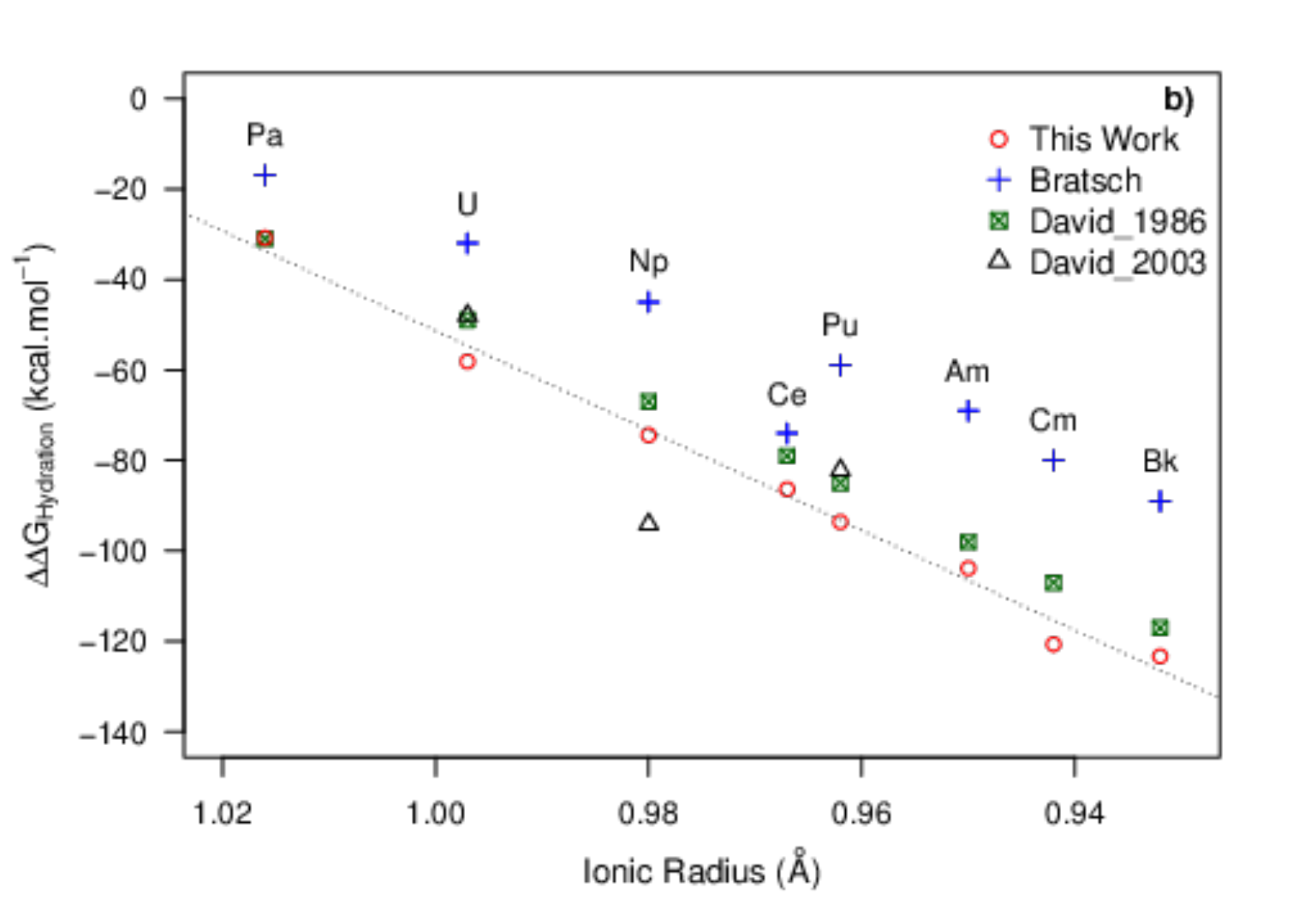}
\caption {(a) Distances $d_{AnO}$ (red circles) as a function of ionic radii~\cite{radii-shannon-Acta-Cryst1976-32-751,actinide-David-JLCM1986-121-27}. The $N_c$ estimates, computed by integrating the first RDF peak shown as blue squares in Figure \ref{fig:RDF_An-O_MDfit}, are reported on that plot (as the RDFs are null between their first and second peak the $N_c$ error bar is zero).(b) \ce{An^{IV}} hydration free energy relative to \ce{Th^{IV}} ($\Delta \Delta G_{hyd}$) versus the \ce{An^{IV}} ionic radii.  In dashed lines, the linear regression fit, the regression coefficients are 0.99 and 0.99 for (a) and (b) respectively. Experimental data taken from the work  by David~{\etal}~\cite{actinide-David-JLCM1986-121-27,actinide-David-NJC2003-27-1627} and Bratsch and Lagowski~\cite{actinide-Bratsch-JPC1986-90-307}.}
\label{fig:dAn-O_MDfit_vs_ir}
\end{figure}

As expected from the \ce{An^{IV}} large positive charge, MD simulations show the tetravalent cation surrounding water molecules to be organized in two well defined solvation shells (see Figure~\ref{fig:RDF_An-O_MDfit}). At the exception of \ce{Cm^{IV}}, the RDF first peak shifts in a very progressive fashion to shorter distances as the actinide gets heavier. The $d_{AnO}$ distance decreases from \ce{Th^{IV}} to \ce{Cm^{IV}}, while the \ce{Bk^{IV}} one is found to be slightly longer than for \ce{Cm^{IV}}. Our $d_{AnO}$ distance agree within at most \SI{0.02}{\angstrom} compared to available EXAFS data (see Table~\ref{tab:MD_CN_Rint}).  We plot on Figure~\ref{fig:dAn-O_MDfit_vs_ir} the $d_{AnO}$ distances as a function of the \ce{An^{IV}} ionic radii estimated from crystallographic data~\cite{radii-shannon-Acta-Cryst1976-32-751,actinide-David-JLCM1986-121-27}. The two series of data are linearly correlated. Following the idea of Warren and Patel~\cite{md-Warren-JCP2007-127-064509}, the ionic radii can be estimated from the distances $d_{AnO}$ to which the water molecular radius is subtracted. For the TCPE/2013 model, the water radius is \SI{1.393}{\angstrom}~\cite{actinide-Real-JCP2013-139-114502}. That yields a nice agreement between our computed radii and the crystallographic ones, within less than \SI{0.02}{\angstrom} (see Table S3 in the ESI).

%FIGURE dAn-O fct du rayon ionique
%FIGURE dAn-O fct du rayon ionique

%%%%%%%%%%%%%%%%%%%%%%%% AVEC MD litterature
%SI
%TABLE DES DISTANCES ET NC 
%%%%%%%%%%%%%%%%%%%%%%
\begin{table*}[ht]
\centering
%\tiny
\caption{Comparison of structural MD computed data for the series of \ce{An^{IV}-{water}} with GP-P and BP-P to experimental EXAFS data and other MD simulations.} 
\begin{tabular}{lrlrlrlcrlc}
\toprule
& \multicolumn{2}{c}{\textit{GP-P}} &  \multicolumn{2}{c}{\textit{BP-P}}&\multicolumn{3}{c}{Other MD simulations} &  \multicolumn{3}{c}{Experiment} \\
\midrule
Elt & CN & d$_{An-O}$  (\AA)& CN &d$_{An-O}$  (\AA)& CN &d$_{An-O}$  (\AA)& Refs. & CN &d$_{An-O}$  (\AA)  & Refs.\\
\midrule
%Th &10.0&2.48&9.4& 2.48 & 9, 9, 8.05-8.45, 8-9, 9.5-11.4, 8-9    & 2.49, 2.4, 2.44-2.47, 2.48, 2.45-2.47, 2.46-2.49    &   \cite{actinide-Atta-Fynn-JPC2016-120-10216,actinide-Marjolin-TCA2012-131-1,actinide-Real-JPC2010-114-15913,actinide-Real-JCC2013-34-707,actinide-Spezia-JPC2012-116-6465}   &9.0 - 12.7&2.44 - 2.46&\cite{actinide-Moll-IC1999-38-1795, actinide-Torapava-IC2009-48-11712, actinide-Hennig-IC2007-46-5882, actinide-Neck-RA2002-90-485, actinide-Wilson-ACIE2007-46-8043}\\
Th &10.0&2.48&9.4& 2.48 & 8 - 11.4  & 2.40 - 2.50    &   \cite{actinide-Atta-Fynn-JPC2016-120-10216,actinide-Marjolin-TCA2012-131-1,actinide-Real-JPC2010-114-15913,actinide-Real-JCC2013-34-707,actinide-Spezia-JPC2012-116-6465}   &9.0 - 12.7&2.44 - 2.46&\cite{actinide-Moll-IC1999-38-1795, actinide-Torapava-IC2009-48-11712, actinide-Hennig-IC2007-46-5882, actinide-Neck-RA2002-90-485, actinide-Wilson-ACIE2007-46-8043}\\
Pa  &10.0&2.44 &9.1 &2.45 &-&-&- &-&2.43&\cite{actinide-Banik-DT2016-45-453}\\
%U    &9.9&2.41&9.0&2.42 &8.7, 9&2.45, 2.45&\cite{actinide-Atta-Fynn-IC2012-51-3016,actinide-Frick-IC2009-48-3993}&8.7 - 10.6&2.40 - 2.42&\cite{actinide-Moll-IC1999-38-1795,actinide-Ikeda-Ohno-IC2009-48-7201,actinide-Hennig-IC2005-44-6655}\\
U    &9.9&2.41&9.0&2.42 &8.7 - 9&2.45&\cite{actinide-Atta-Fynn-IC2012-51-3016,actinide-Frick-IC2009-48-3993}&8.7 - 10.6&2.40 - 2.42&\cite{actinide-Moll-IC1999-38-1795,actinide-Ikeda-Ohno-IC2009-48-7201,actinide-Hennig-IC2005-44-6655}\\
Np  &9.9&2.40&9.0&2.40&-&-&-&8.7 - 10.4&2.37 - 2.40&\cite{actinide-Antonio-RA2001-89-17,actinide-Ikeda-Ohno-IC2008-47-8294,Denecke-Tal2005-65-1008}\\
Pu  &10.0&2.39&9.0&2.38&8.1&2.41&\cite{actinide-Odoh-JPC2013-117-12256} &7.8 -~ ~8.4&2.38 - 2.39&\cite{actinide-Rothe-IC2004-43-4708,actinide-Dardenne-RA2009-97-91}\\
Am &9.0&2.33&8.7&2.35&-&-&-&- &-&-\\
Cm &8.9&2.32&8.3&2.335&-&-&-&-&-&-\\
Bk  &9.5&2.34&8.9&2.34&-&-&-&7.9&2.32&\cite{actinide-Antonio-RA2002-90-851}\\
\midrule
Ce &9.1&2.365&8.8&2.38&9&2.44&\cite{lanthanide-lutz-IC2012-51-6746}&< 9&2.42&\cite{lanthanide-Sham-PRB1989-40-6045}\\
\bottomrule
\end{tabular}
\label{tab:MD_CN_Rint}
\end{table*}

%TABLE DES DISTANCES ET NC 
%%%%%%%%%%%%%%%%%%%%%%%% AVEC MD litterature
%%%%%%%%%%%%%%%%%%%%%%%%%%%

% Nc

The computed $N_c$'s start at 9.4 for \ce{Th} and decrease until they reach a plateau of about 9 for the series \ce{Pa-Pu, Bk}\ with \ce{Am} and \ce{Cm} moving down away from the series with a coordination number of respectively 8.7 and 8.3, see Figure~\ref{fig:dAn-O_MDfit_vs_ir}. Compared to experimental data regarding solutions for which the counter ions are known to not enter the \ce{An^{IV}} first hydration shell (i.e. mainly perchlorate anions), $N_c$'s for the \{\ce{Th^{IV}} - \ce{Np^{IV}}\} series are included within the experimental range, while the $N_c$ values for \ce{Pu^{IV}} and \ce{Bk^{IV}} are one unit above the experimental values. However, the EXAFS uncertainty regarding $N_c$ values being \SIrange{10}{20}{\percent}, our $N_c$ values for \ce{Pu^{IV}} and \ce{Bk^{IV}}  may be considered as agreeing with experiment, within the error bars. 

%Coordination mode

%FIGURE geom Polys
\begin{figure}[ht]
%\begin{figure}
\centering
\begin{subfigure}[t]{0.45\linewidth}
\centering
\includegraphics[width=0.5\linewidth]{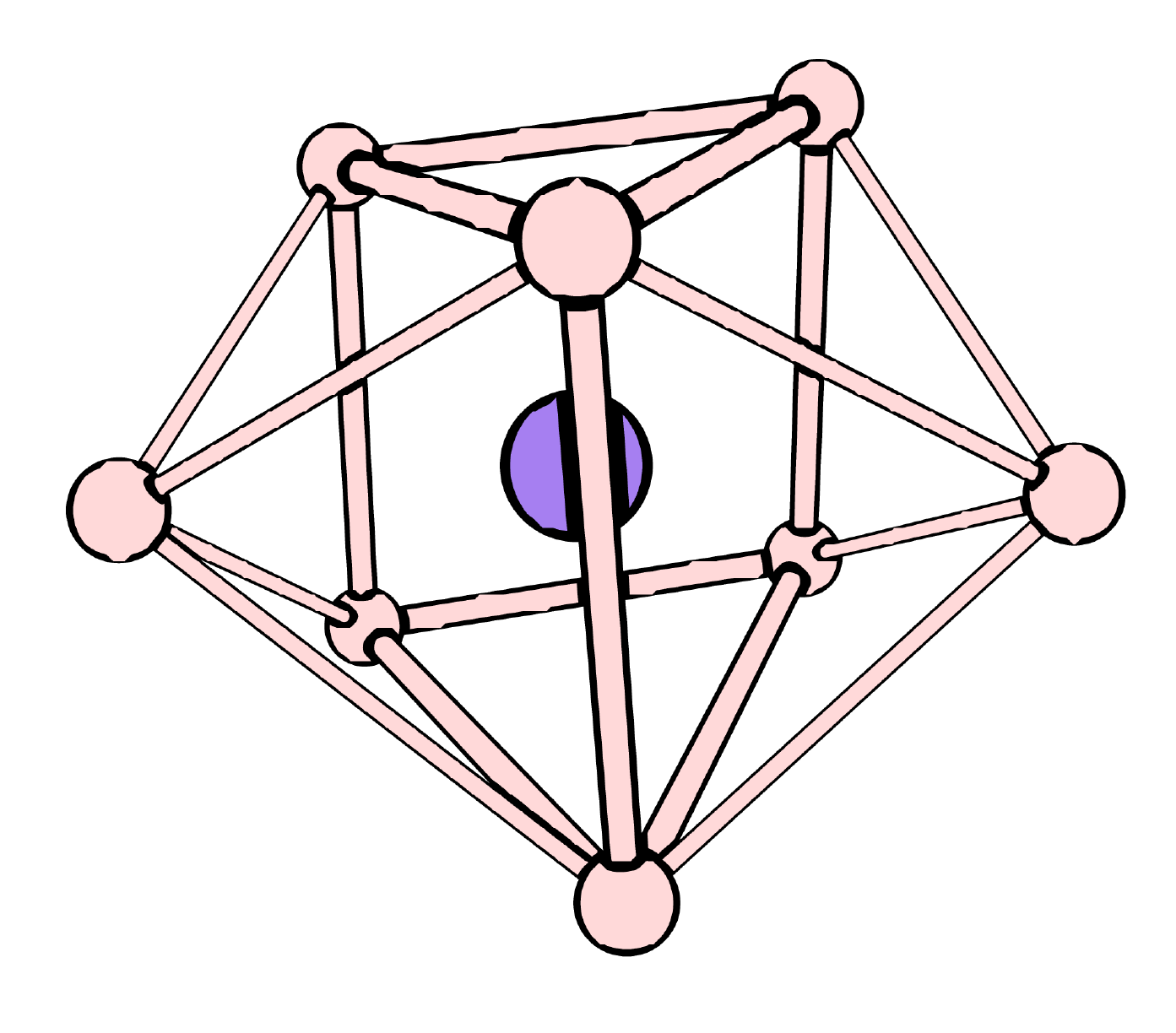}
\caption{Biaugmented Trigonal Prism (CN=8)}
\end{subfigure}
\hfill
\begin{subfigure}[t]{0.45\linewidth}
\centering
\includegraphics[width=0.5\linewidth]{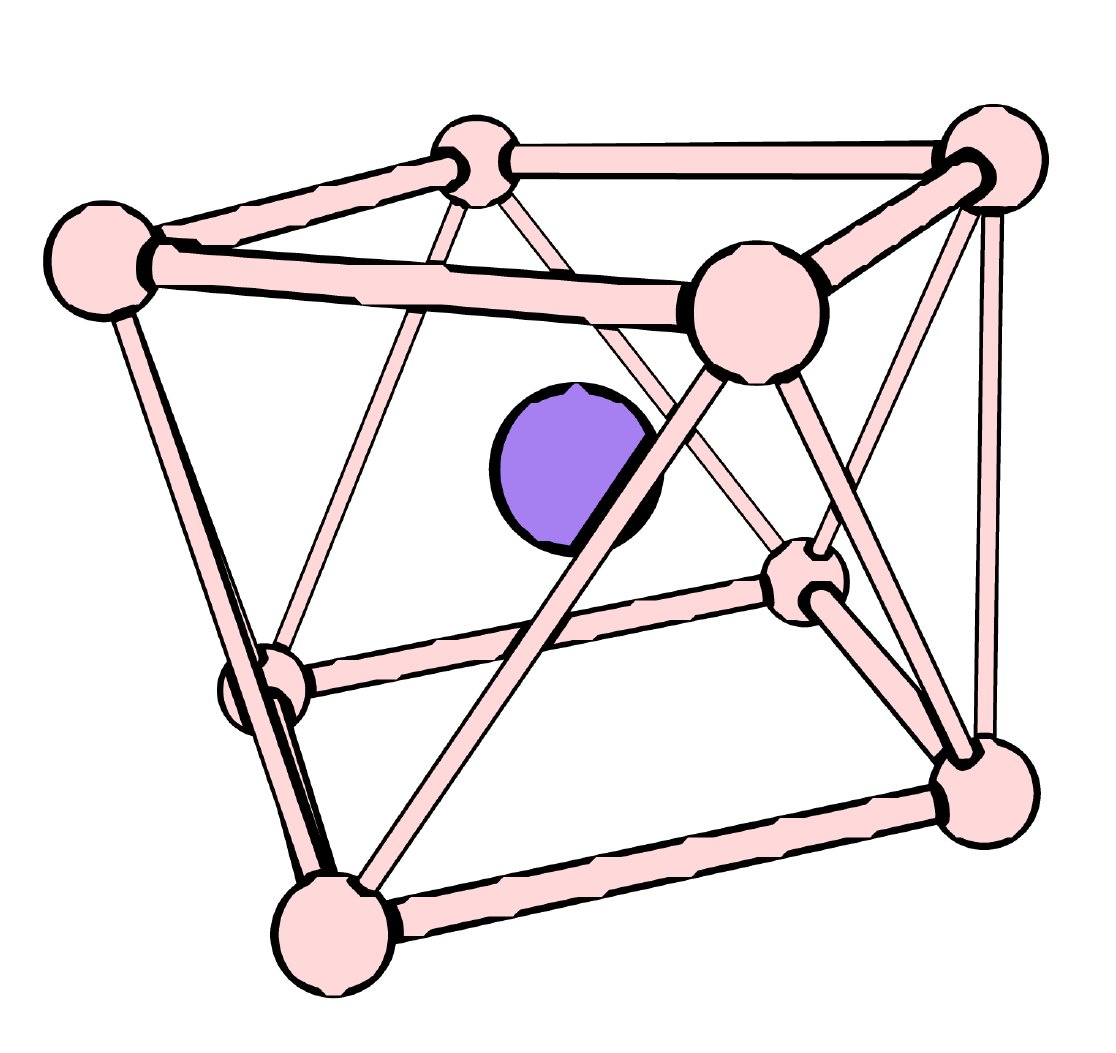}
\caption{Square Antiprism (CN=8)}
\end{subfigure}

\begin{subfigure}[t]{0.45\linewidth}
\centering
\includegraphics[width=0.5\linewidth]{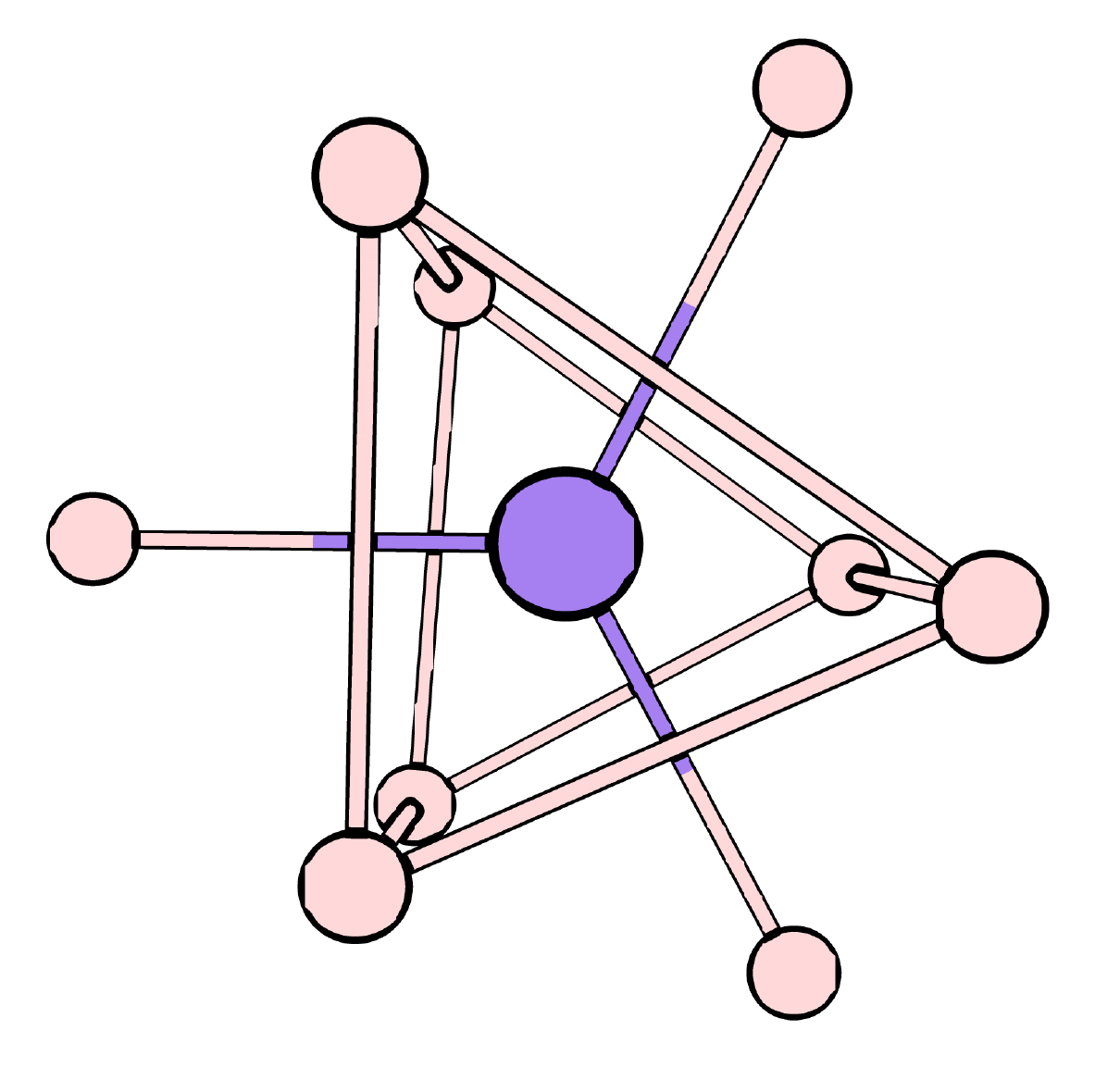}
\caption{Triaugmented Trigonal Prism (CN=9)}
\end{subfigure}
\hfill
\begin{subfigure}[t]{0.45\linewidth}
\centering
\includegraphics[width=0.5\linewidth]{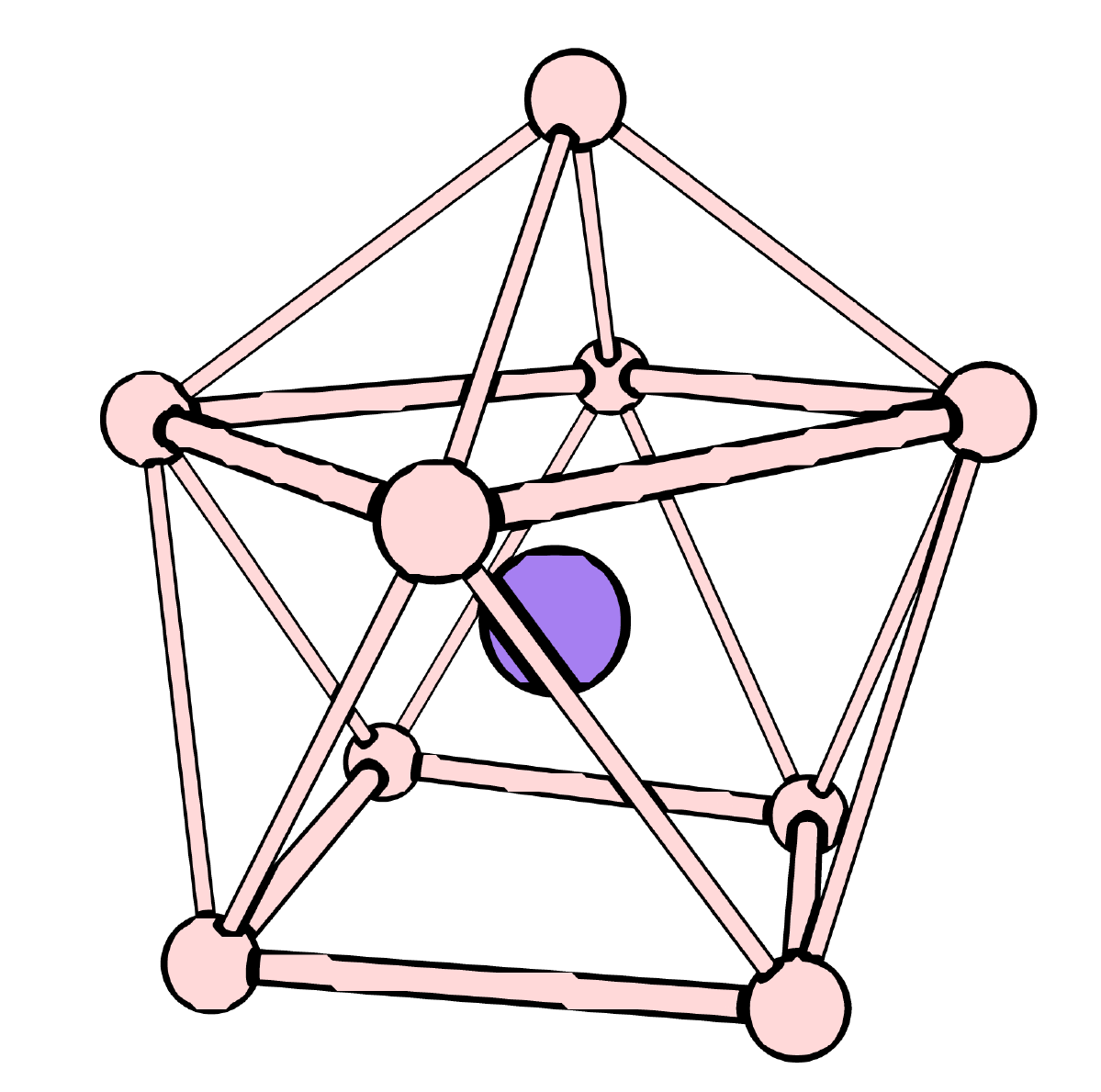}
\caption{Monocapped Square Antiprism (CN=9)}
\end{subfigure}

\begin{subfigure}[t]{0.45\linewidth}
\centering
\includegraphics[width=0.5\linewidth]{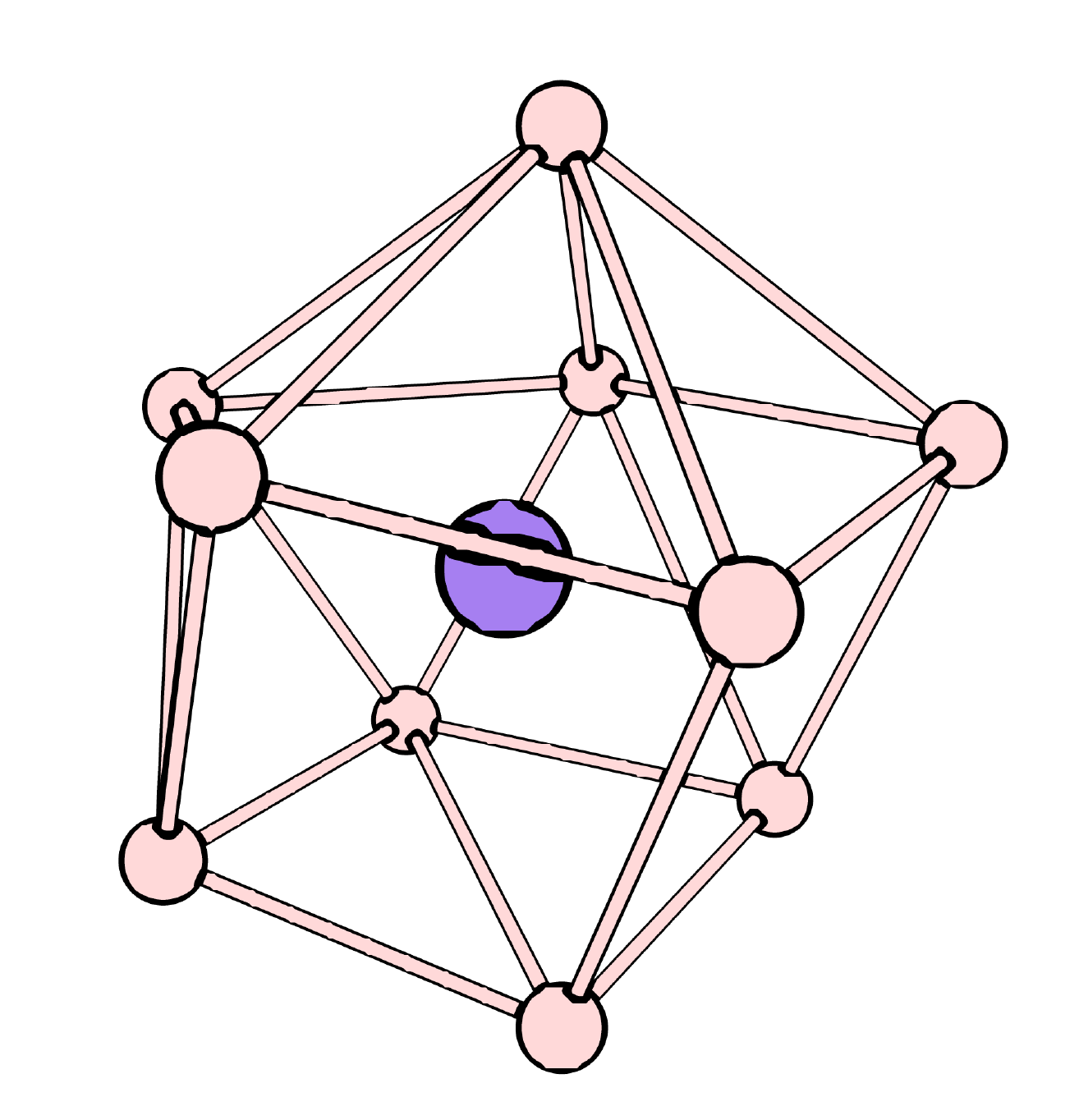}
\caption{Sphenocorona (CN=10)}
\end{subfigure}
\hfill
\begin{subfigure}[t]{0.45\linewidth}
\centering
\includegraphics[width=0.5\linewidth]{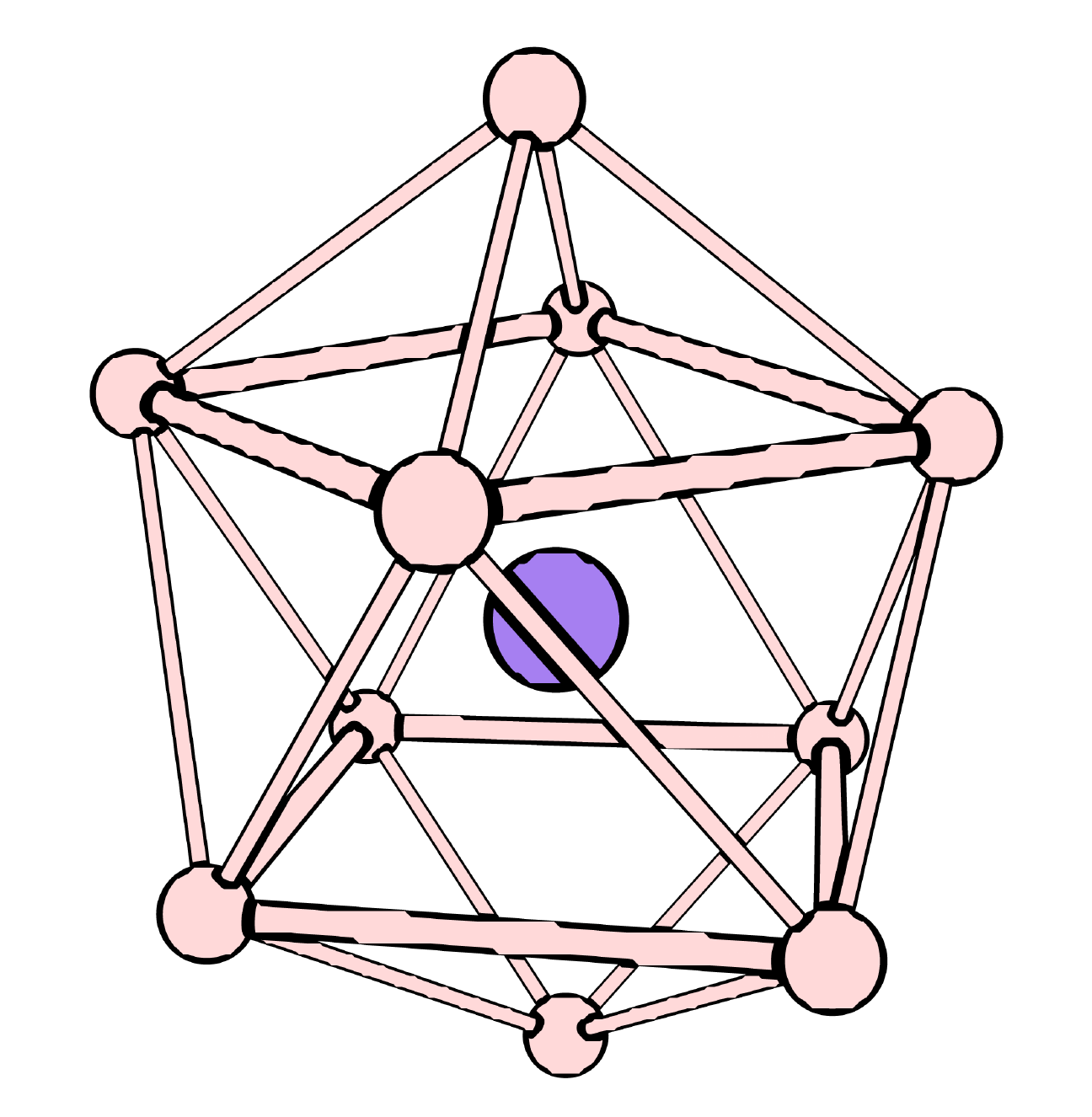}
\caption{Bicapped Square Antiprism (CN=10)}
\end{subfigure}

\caption{Observed polyhedra for the coordination numbers (CN) ranging from 8 to 10.} 
\label{fig:polyhedra}
%\end{figure}
\end{figure}
%FIGURE geom Polys

The fractional $N_c$'s for  \ce{Th^{IV}}, \ce{Pa^{IV}}, \ce{Am^{IV}}, \ce{Cm^{IV}} and \ce{Bk^{IV}}, reveal the existence of an equilibrium between two coordination modes. An analysis of the coordination polyhedra  has been made with \textit{ChemNetworks}~\cite{ChemNetwoks-Ozkanlar-JCC2014-35-495} and shows that the 8 coordination mode is  associated with the formation of either a square anti-prismatic or a biaugmented trigonal prismatic geometry of the water molecules around the actinide cation. The identified polyhedra for the 9 coordination number are either a triaugmented trigonal prism or a mono-capped square anti-prism and for the 10 coordination number either a bicapped square anti-prism or a sphenocorona. The corresponding structures for the observed polyhedra extracted from the MD trajectories can be found in Figure~\ref{fig:polyhedra}. These coordination modes are in line with the ones observed for some studied \ce{An^{IV}}, in particular for \ce{Th^{IV}} and \ce{U^{IV}}~\cite{actinide-Atta-Fynn-IC2012-51-3016, actinide-Atta-Fynn-JPC2016-120-10216, actinide-Spezia-PCCP2014-16-5824}.

%MRT

We computed the mean residence time, MRT, of a water molecule within \ce{An^{IV}} first hydration shell along the MD trajectories from the approach of Impey~{\etal}~\cite{MRT-Impey-JPC1983-87-5071}. The computed MRTs (See Table~S2 in the ESI) range from a few hundreds of picoseconds for \ce{Cm^{IV}}, \ce{Am^{IV}} and \ce{Th^{IV} } up to \SI{1.5}{ns} for \ce{U^{IV}}. Experimentally, only the \ce{U^{IV}} MRT has been reported~\cite{MRT-Banyai-JPCA2000-104-1201}. about \SI{185}{\ns}, as well as an upper bound limit for \ce{Th^{IV}}: \SI{20}{\ns}~\cite{MRT-Banyai-JPCA2000-104-1201}. Our estimates are \emph{a priori} from one to two orders of magnitude smaller than experiment. This discrepancy may come from some deficiencies of our model to reproduce accurately the energy barriers in the water exchange reaction. On the other hand, we simulated here a single actinide cation surrounded by pure water whereas, in the experiment, very salty solutions are used (around \SI{4}{\Molar} of perchlorate). The presence of concentrated salts and counter-ions may also alter the cation hydration structure and affect the water dynamics at its vicinity~\cite{actinide-Atta-Fynn-JPC2016-120-10216}. The quality of the present BP-P force field cannot be incriminated yet to explain the apparent disagreement between the computed and the rare experimental MRT estimates and further investigations should be carried on.

% the variations being correlated with the existence of an equilibrium between two coordination numbers
% with a $\tau^{*}$  of 2 ps. 
%Discussion par

%DG hydratation

We computed the change in hydration free energies for the \ce{An^{IV}} series relative to \ce{Th^{IV}} ($\Delta\Delta G_{hyd}$)  using the thermodynamic integration method. As for the distances $d_{AnO}$, we obtain a linear correlation between our $\Delta\Delta G_{hyd}$  values and the crystallographic ionic radii (see Figure \ref{fig:dAn-O_MDfit_vs_ir}).
% as well as between our  $\Delta G_{hyd}$  values and the $\Delta G_{hyd}$ ones computed from the David~{\etal}'s model~\cite{actinide-David-JLCM1986-121-27,actinide-David-NJC2003-27-1627} (in the latter case, at the exception of \ce{Np^{IV}}, see our discussion in Introduction), see Figure \ref{fig:dAn-O_MDfit_vs_ir}.  
Our purely ab initio approach (BP-P force fields) predicts the $\Delta\Delta G_{hyd}$ values to smoothly decrease along the \ce{An^{IV}} series. This is in line with the expectation that the continuous decrease of the $d_{AnO}$ distances along that series yields a reinforcement of \ce{An^{IV}}/water interactions. These results compare very well to David~{\etal}'s estimates~\cite{actinide-David-JLCM1986-121-27,actinide-David-NJC2003-27-1627} which all exhibit a decreasing trend much steeper than the one reported by Bratsch and Lagowski~\cite{actinide-Bratsch-JPC1986-90-307}. The discrepancies between the empirical models question the available data reliability and prove the challenges addressed to ab initio modelling approaches. 
%It seems that David's approach, based on a physical model 
%rather than an empirically fitted trend with rare experimental data \cite{lanthanide-Bratsch-JPC1985-89-3310,lanthanide-Bratsch-JPC1985-89-3317}

We also assigned BP-P parameters from the iterative procedure for the lanthanide element \ce{Ce^{IV}}. We simulated then that element in an aqueous solution using the same computational protocol and simulation conditions as above. From our simulations, \ce{Ce^{IV}} exhibits a very similar behavior to \ce{Pu^{IV} } in bulk water, with an identical $d_{AnO}$ distance of 2.38~{\AA} and close $N_c$ values, of respectively 9.0 and 8.8, for \ce{Pu^{IV}} and \ce{Ce^{IV}}. The similarity is further consolidated while comparing the respective RDFs of both elements that can be almost superimposed. Lastly, concerning the $\Delta G_{hyd}$ values, the \ce{Ce^{IV}} estimate is found to be in between \ce{Np^{IV}} and \ce{Pu^{IV}}, still closer to the latter with a difference of \SI{8}{\kcal}. Such a level of similarities between hydrated \ce{Ce^{IV}} and  \ce{Pu^{IV}} have been also reported from QM and experimental studies~\cite{actinide-Sulka-JPCA2014-118-10073,lanthanide-Marsac-DaltonTrans2017-46-13553}.

Finally, considering the previous MD simulations reported in the literature, on the one hand, we find out that DFT-based MD generally underestimate the coordination number~\cite{actinide-Atta-Fynn-IC2012-51-3016}, while exhibiting the largest computed interaction distances (Table~\ref{tab:MD_CN_Rint}); while on the other hand, different classical MD for \ce{Th^{IV}} give very large ranges of coordination numbers (from 8 to 11.4) and interaction distances (from \SIrange{2.40}{2.48}{\angstrom}). This disparity further proves the need for highly accurate force-field parameters assignment protocols to reduce as far as possible all the drawbacks mentioned in Introduction about building force fields.

\section*{Conclusion}
In conclusion, we report the first polarizable \ce{An^{IV}-H2O} classical potentials derived from a self-consistent iterative \textit{ab initio} methodology for the series from thorium to berkelium. We emphasize that the specific inclusion of sampled condensed phase reference structures in a self-consistent manner enables to reach an adequate accuracy necessary to simulate the actinide aqueous condensed phase systems. The hydration study from classical molecular dynamics with our newly developed force field reveals that, except for \ce{Am^{IV}} and \ce{Cm^{IV}} which sit aside the series, the structural and thermodynamical properties change in a very smooth fashion along the studied series. The first comparisons to experimental data exhibit a very good overall agreement. 
The increase of the hydration energies along the tetravalent actinide series has some impact on their geomobility. Indeed, one might expect all An(IV) to form like Th(IV) outer-sphere interactions with muscovite-like minerals~\cite{actinide-Geckeis-CR2013-113-1016}, while strongly complexing surfaces with carbonate, phosphate, or carboxylic acid may form inner-sphere complexes, calling for further extensions of force-field interaction models to such anionic ligands. Thus, this study will be extended in a near future by considering counter anions to achieve a thorougher comparability to experimental conditions. 
%Nevertheless, these methodological developments pave the way for a better classical description of the +4 actinides in solution, and thus our ability to study their chemical behavior in more complex and chemically relevant systems; this being of special interest considering how important and challenging \ce{Pu^{IV}} chemistry is with respect to the nuclear fuel cycle.
Finally, this results and methodological developments made here for the most highly charged ions of the periodic table are strictly transferrable to any ion or cation in solution for which an exploration of several properties in water or other solvent is envisaged, being either related to biological or atmospheric issues.  

\section*{Conflicts of interest}
There are no conflicts to declare.

\section*{Acknowledgements}
We acknowledge support by the French government through the Program "Investissement d'avenir"  through the Labex CaPPA (contract ANR-11-LABX-0005-01) and I-SITE ULNE project OVERSEE (contract ANR-16-IDEX-0004), CPER CLIMIBIO (European Regional Development Fund, Hauts de France council, French Ministry of Higher Education and Research), French national supercomputing facilities (grants DARI x2016081859 and A0050801859).

\section*{Supplementary material}
The supplementary material includes:
\begin{itemize}
  \item Discussion of the accuracy of DFT for interaction energies
  \item Parameters of the force-field model for ion/water interactions
  \item Convergence of the BP-P process.
\end{itemize}

\bibliography{MD-AnIV} %You need to replace "rsc" on this line with the name of your .bib file

\end{document}

\clearpage
\tableofcontents

\clearpage
\listoftables

\clearpage
\listoffigures
\clearpage

\section{Comments on the accuracy of DFT for total binding energies and fragment interaction energies}
\label{sec:DFT_err}
\begin{figure}
\centering
\includegraphics[width=\linewidth]{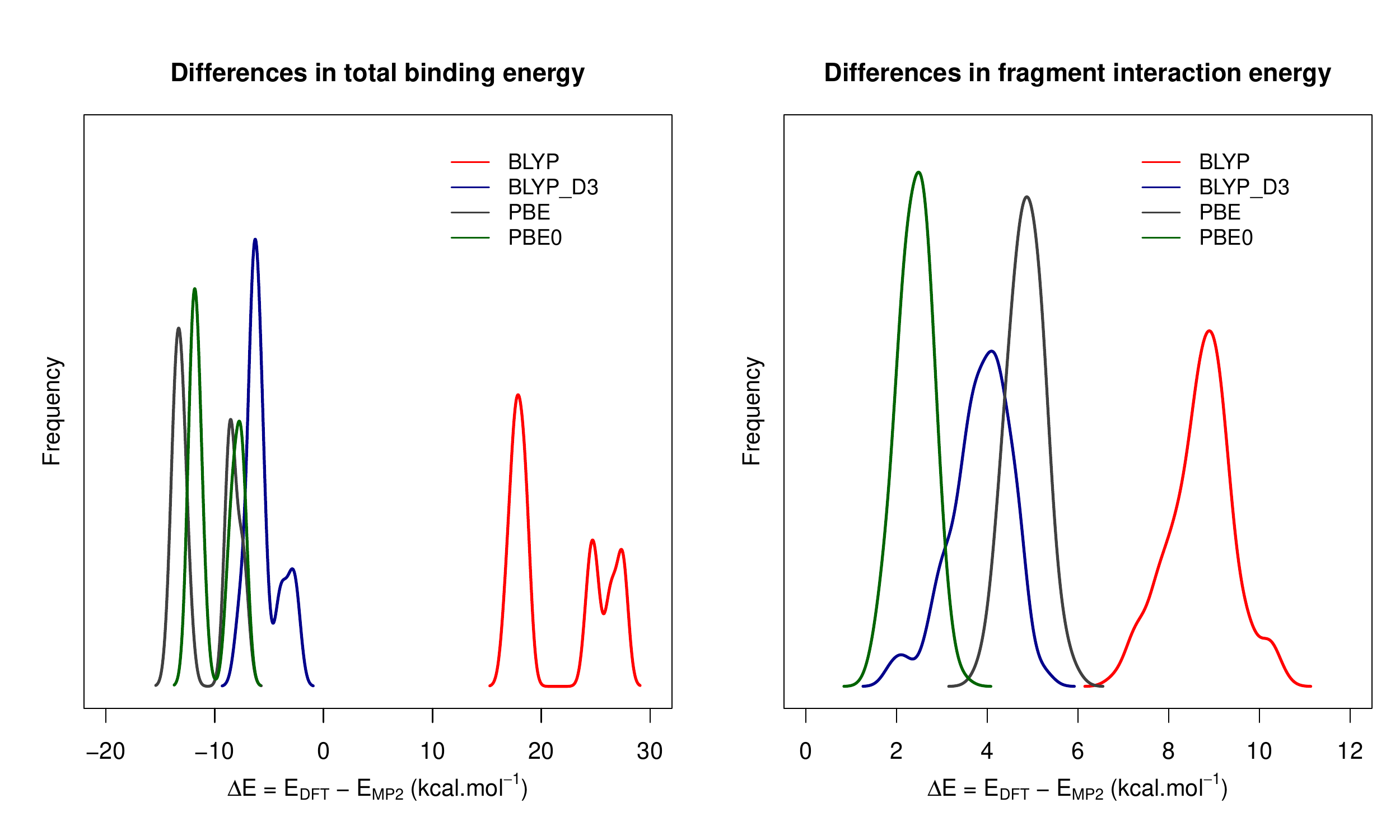}
\caption{Differences between DFT and MP2 energies calculated on twenty clusters extracted from \ce{Th^{IV}} MD for (left) total binding energy and (right) fragment interaction energies. For the total binding energies, the double peaks correspond to different coordination numbers (\ce{Th^{IV}(H2O)_10} and \ce{Th^{IV}(H2O)_9} here).} 
\label{fig:DFT_err}
\end{figure}

In the present study, we chose to compute all cluster binding energies and fragment interaction energies at the correlated MP2 level of theory, but one may question the accuracy of available functionals of the density (DFT), which depends on the simultaneous accuracies of actinide-water and water-water interactions. Réal~{\etal}\cite{actinide-Real-JPC2010-114-15913} showed that GGA (BP86), a hybrid (B3LYP), and meta-GGA (M06) functionals all overestimate the Th(IV)-water interaction energy by up to \SI{18}{\kcal}. Furthermore, most functionals exhibit excessive repulsions.\cite{water-Gillan-JCP2016-144-130901}
In Figure~\ref{fig:DFT_err}, we have compared the total binding energies and fragment interaction energies (interaction energy between a water molecule and the \ce{An^{IV}(H2O)_{n-1} cluster}, computed with two GGA functionals, BLYP, PBE, one hybrid functional, PBE0, and the dispersion corrected BLYP+D3 one. All four functionals of the density overestimate the fragment interaction energies, confirming the fact that water-water repulsions are overestimated. This error is not counterbalanced by the bias in the metal-water interactions; while BLYP overestimates total binding energies, PBE, PBE0, and BLYP+D3 yield to negative deviations, the larger the hydration number the larger the deviation from MP2 energies. This confirms that currently available functionals are not accurate enough to be used for the calculations of the QM reference energies.

%FIGURE Erreurs DFT
%\begin{figure*}

\section{BP-P convergence}
\label{sec:iterative-fit}
\begin{figure*}
%\begin{figure}
\centering
\begin{subfigure}[t]{\linewidth}
\centering
\includegraphics[width=\linewidth]{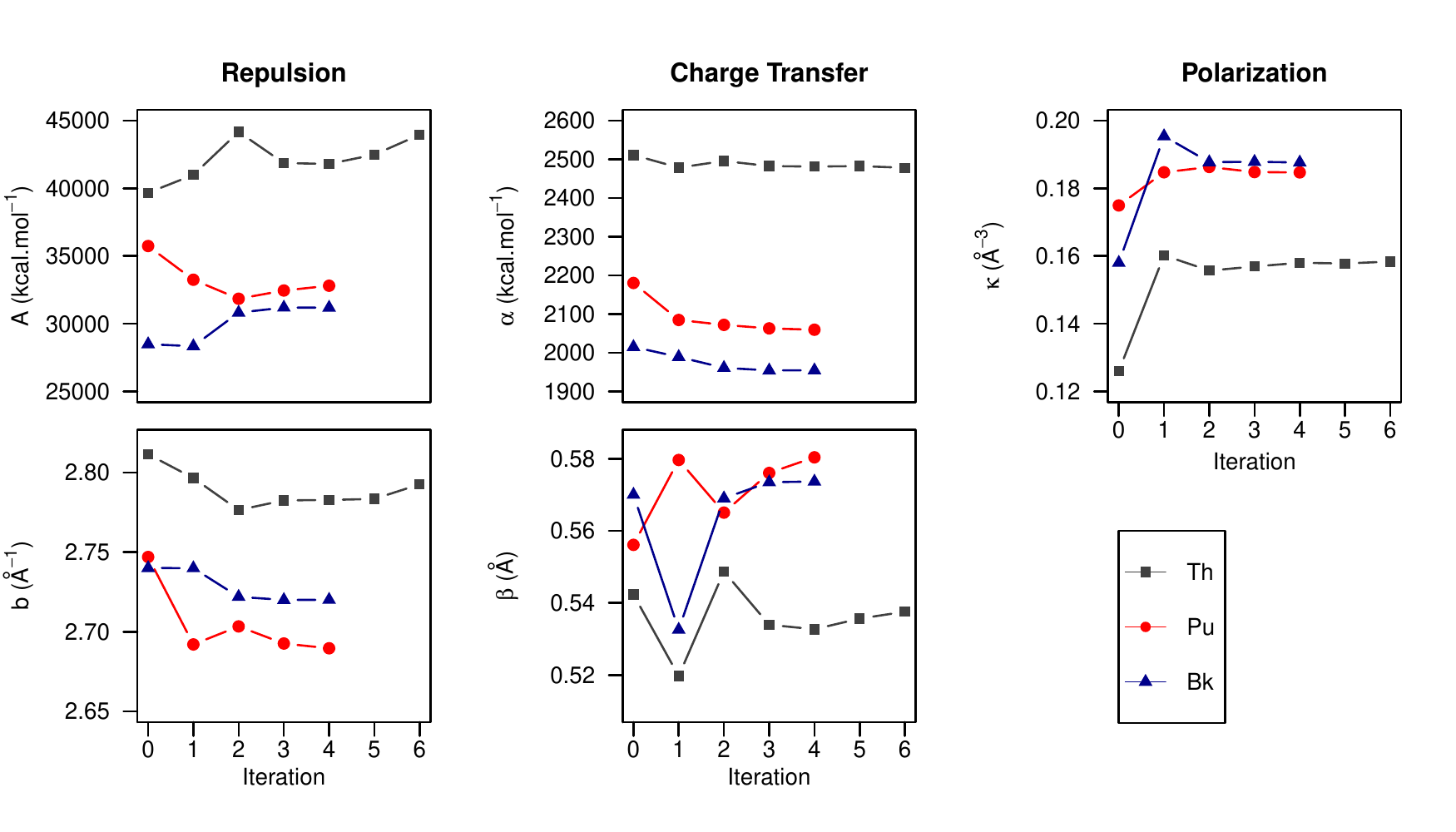}
\caption{Force-field parameters.}
\end{subfigure}

\begin{subfigure}[t]{0.63\linewidth}
\centering
\includegraphics[width=\linewidth]{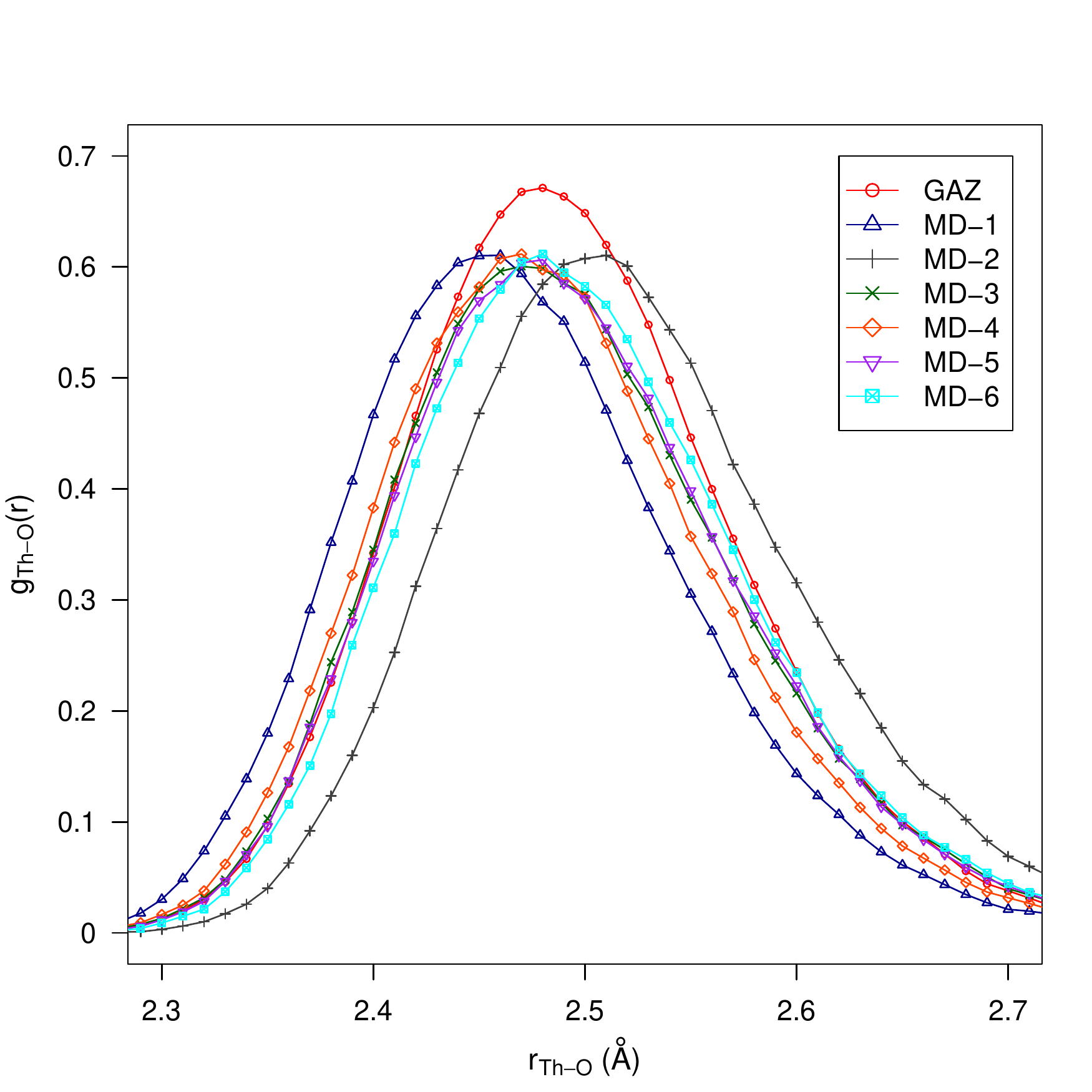}
\caption{\ce{Th^{IV}-O} RDF.}
\end{subfigure}
\hfill
\begin{subfigure}[t]{0.33\linewidth}
\centering
\includegraphics[width=\linewidth]{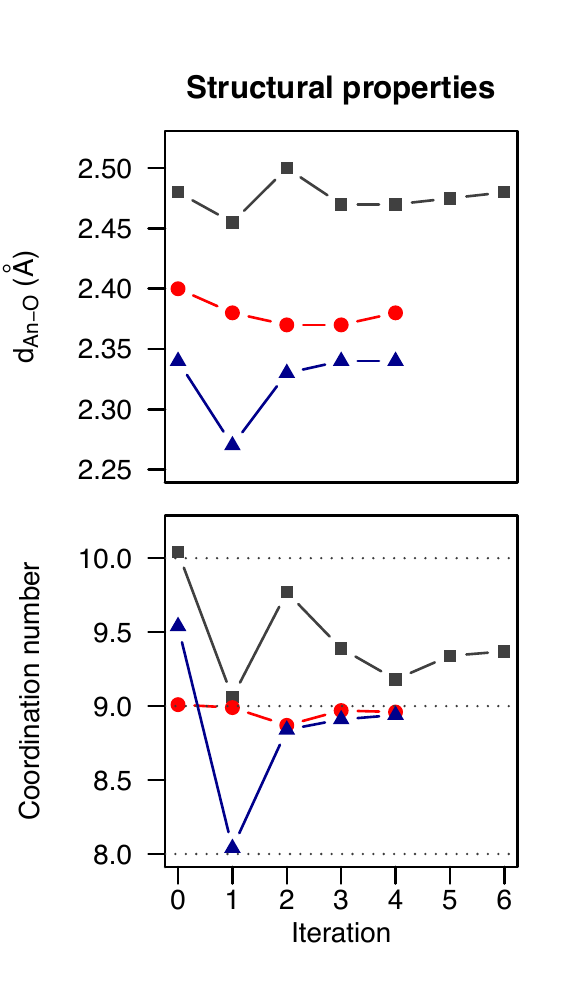}
\caption{Structural properties.}
%\caption{CN and $d_{\ce{An^{IV}-H2O}}$}
\end{subfigure}

\caption{(a)~Convergence of~FF parameters. (b)~Evolution of the \ce{Th^{IV}-O} radial distribution function (RDF) along the iterations of the sampling of the solvated phase procedure \textit{BP-P}. (c)~Convergence of~coordination number and \ce{An^{IV}-H2O} interaction distances for \ce{Th^{IV}}, \ce{Pu^{IV}} and \ce{Bk^{IV}}.  .} 
\label{fig:FF_convergence}
%\end{figure}
\end{figure*}

Figure~\ref{fig:FF_convergence} (a) illustrates that the five FF parameters for Th, Pu, and Bk converge after 3-5 iterations, as the training data sets is updated with sampled snapshots from classical MD simulations. The convergence of the \ce{An-H2O}  interaction distances and radial distribution functions as well as average coordination numbers can also be visualized on Figure~\ref{fig:FF_convergence} subsets (b) and (c).

%%%%%%%%%%%%%%%%%%%%%%%%%
%FIGURE CONVEGENCE
%FIGURE CONVERGENCE
%%%%%%%%%%%%%%%%%%%%%%%%%%

%%%%%%%%%%%%%%%%%%%%%%%%%%%
% SI PARAMETRES FINAUX
%%%%%%%%%%%%%%%%%%%%%%%%%%%

\section{Final parameters}
%TABLE PARAMETRES MD-FIT
\begin{table}[htp]
%\begin{table}%[tbhp]
\centering
\caption{Force-field parameters for the series of \ce{An^{IV}-H2O} potentials.}
\begin{tabular}{crcccccc}
\toprule
 && $A$ & $B$ & $D^{ct}$ & $\beta$ & $\kappa$ &$\alpha$ \\
&Element&(\si{\kcal})&(\si{\per\angstrom})&(\si{\kcal})&(\si{\angstrom})&(\si{\per\cubic\angstrom})&(\si{\cubic\angstrom})\\
\midrule
%%%%% Method 1 table
 & & & & & & & \\
\multirow{9}{*}{\rotatebox[origin=c]{90}{GP-P}} & Th &39673&2.81&2511&0.54&0.126 &1.142\\
&Pa &33783&2.75&2197&0.56&0.140 &1.217\\
&U &30391&2.72&2034&0.57&0.148&1.180\\
&Np &26456&2.68&1715&0.59&0.146&1.063\\
&Pu &30692&2.76&2027&0.57&0.148&1.063\\
&Am &20823&2.57&1060&0.67&0.178&1.000\\
&Cm &19251&2.55&770&0.73&0.180&1.000\\
&Bk &28490&2.74&2015&0.57&0.158&1.000\\
&Ce & 22081 & 2.60 & 1074 & 0.66 & 0.164 &0.860\\
%%%%%% End method 1 part
 & & & & & & & \\
\midrule
 & & & & & & & \\
%%%%%% Begining of method2
\multirow{9}{*}{\rotatebox[origin=c]{90}{BP-P}} &Th &43960&2.792&2478&0.538&0.158&1.142 \\
&Pa &42997&2.756&2396&0.559&0.176 &1.217\\
&U &36135&2.704&2395&0.561&0.179&1.180\\
&Np &36402&2.725&2208&0.568&0.182&1.116\\
&Pu &32808&2.690&2060&0.580&0.185&1.063\\
&Am &50976&2.930&2108&0.543&0.208&1.000\\
&Cm &35532&2.707&2015&0.590&0.215&1.000\\
&Bk &31187&2.720&1955&0.574&0.188&1.000\\
%\cline{2-8}
&Ce & 42891 & 2.867 & 1348 & 0.591& 0.197&0.860\\
 & & & & & & & \\
\bottomrule
%%%%%End of method 2

\end{tabular}
\label{tab:FF_MD_parameters}
%\end{table}
\end{table}

%TABLE PARAMETRES 

The actinide charges for Coulombic and polarization are fixed to their +4 net charge and their polarizability is either fixed to the QM values reported in the literature~\cite{actinide-Parmar-JPC2013-117-11874,actinide-Real-PR2008-78-052502} or fixed to \SI{1}{\cubic\angstrom}, the latter assumption not exhibiting any significant impact on the interaction energies. Considering that the water potential parameters are fixed, only five parameters need to be adjusted with the Model-Independent Parameter Estimation (PEST) software package~\cite{prog-pest}; namely the repulsion $A_{An-i}$ and $B_{An-i}$, the charge transfer $D^{ct}_{An-i}$ and $\beta_{An-i}$ and the Thole damping one $\kappa$. The final parameters are listed in Table~\ref{tab:FF_MD_parameters}.

%%%%%%%%%%%%%%%%%%%%%%%%%%%%%%%%%%
% SI POLYEDRES
%%%%%%%%%%%%%%%%%%%%%%%%%%%%%%%%%%

%TABLE MRT 
\begin{table}%[tbhp]
\centering
\caption{Mean Residence Time ($\tau$) in nanosecond of a water molecule in the first coordination sphere of the \ce{An^{IV}}.}
\begin{tabular}{lccccccccc}
\toprule
$\tau$ (\si{\ns}) & Th & Pa & U & Np & Pu & Am & Cm & Bk & Ce  \\
\midrule
MD & 0.6 &0.9&1.5&1.4&1.1&0.5&0.3&1.0&0.5\\
Exp.~\citenum{MRT-Banyai-JPCA2000-104-1201}& < 20 & & $\approx$185 & & & & & & \\
%Exp  & < 20 & & $\approx$185 & & & & & & \\
%Exp~\cite{water-actinide-Ildiko-JPCA2000-104-1201}& < 20 & & $\sim$200 & & & & & & \\

%valeur centrale exp U= 185ns (min=167), (max=208)
\bottomrule
\end{tabular}
\label{tab:MRT}
\end{table}
%TABLE MRT

%\section{Hydration free energies}
%TABLE HYDRATION FREE ENERGIES
\begin{table}
%\begin{table}%[tbhp]
\centering
\caption{Hydration free energies of the \ce{Th^{IV}}-\ce{Bk^{IV}} tetravalent actinide series and \ce{Ce^{IV}} relative to thorium (\si{\kcal}), and ionic radii in \si{\angstrom}.}
\begin{tabular}{lcccccccc | c}
\toprule
  & Pa & U & Np & Pu & Am & Cm & Bk & Ce & Th (absolute) \\
\midrule
 This work & -31 & -58 & -74 & -94 & -104 & -121 & -123 & -86 &  \\
 Refs.~\citenum{lanthanide-Bratsch-JPC1985-89-3310,lanthanide-Bratsch-JPC1985-89-3317,actinide-Bratsch-JPC1986-90-307}& -17 & -32 & -45 & -59 & -69 & -80 & -89 & -74 & -1401 \\
Ref.~\citenum{actinide-David-JLCM1986-121-27}  & -31 & -49 & -67 & -85 & -98 & -107 & -117 & -79 & -1400 \\
 Ref.~\citenum{actinide-David-NJC2003-27-1627}  &  & -48 & -94 & -82 &  &  &  &  & -1457 \\
\midrule
 ionic radii\cite{radii-shannon-Acta-Cryst1976-32-751,actinide-David-JLCM1986-121-27} & 1.016 & 0.997 & 0.980 & 0.962 & 0.950 & 0.942 & 0.932 & 0.967 & 1.048\\
\bottomrule
\end{tabular}
\label{tab:HFE}
\end{table}
%TABLE PARAMETRES 
\clearpage
\bibliography{MD-AnIV}